\documentclass[aps, prd, superscriptaddress, nofootinbib, preprintnumbers]{revtex4}
\usepackage{amsmath}
\usepackage{graphicx}

\newcommand{\lesssim}{\mathrel{\mathpalette\vereq<}}

\newcommand{\chushi}[1]{}

\begin{document}
 \preprint{MISC-2012-07}
 \title{{\bf Discovering 125 GeV techni-dilaton at LHC } 
 \vspace{5mm}}
\author{Shinya Matsuzaki}\thanks{
      {\tt synya@cc.kyoto-su.ac.jp} }
      \affiliation{ Maskawa Institute for Science and Culture, Kyoto Sangyo University, Motoyama, Kamigamo, Kita-Ku, Kyoto 603-8555, Japan.}
\author{{Koichi Yamawaki}} \thanks{
      {\tt yamawaki@kmi.nagoya-u.ac.jp}}
      \affiliation{ Kobayashi-Maskawa Institute for the Origin of Particles and 
the Universe (KMI) \\ 
 Nagoya University, Nagoya 464-8602, Japan.}
\date{\today}

\begin{abstract}

Techni-dilaton (TD) is predicted in walking technicolor (WTC) 
arising as a pseudo Nambu-Goldstone boson 
associated with the approximate scale symmetry spontaneously broken by techni-fermion condensation. 
The TD mass is therefore smaller than those of other techni-hadrons on order of several TeVs, 
small enough to be within reach of the current LHC search. 
 We present a new method to derive the TD couplings directly from the Ward-Takahashi identities, 
which enables us to explicitly calculate the quantities relevant to the TD LHC signatures. 
 To set definite benchmarks, we take one-doublet and 
one-family models of WTC 
and discuss the TD signatures at LHC,  
in comparison with 
those of the standard model (SM) Higgs. 
It is shown that the TD in the one-doublet model 
is invisible at the LHC, while   
the TD signals in the one-family model 
can be found as a large excess  
relative to the SM Higgs 
at around 125 GeV   
only in the diphoton channel.

\end{abstract}
\maketitle

\section{Introduction}

The most urgent issue that the current LHC experiments attempt to settle 
is to clarify a particle responsible for the origin of mass. 
In the standard model (SM) Higgs boson corresponds to the key particle, 
where the mass generation/electroweak (EW) symmetry breaking takes place through an ad hoc assumption of  
the nonzero vacuum expectation value of the elementary Higgs field. 
Thus in the framework of the SM, the origin of mass is put in by hand and not explained,  
which would suggest existence of more fundamental theory beyond the SM.

Recently, the ATLAS~\cite{:2012si,ATLAS-CONF-2012-019} and CMS~\cite{Chatrchyan:2012tx,CMS-PAS-HIG-12-008} 
experiments have excluded  
the SM Higgs boson for most of low mass range up to $\sim$ 600 GeV. 
   If the Higgs-like object, if any, existed above 600 GeV, 
it would be too large to be accounted for by the SM Higgs boson. 
Even for the most recently reported diphoton excess at around 125 GeV~\cite{:2012sk,Chatrchyan:2012tw,CMS-PAS-HIG-12-001},  
the best-fit signal strength denoted by cross section times the $\gamma\gamma$ branching ratio 
is about 2 times larger than that of the SM Higgs resonance~\cite{ATLAS-CONF-2012-019,CMS-PAS-HIG-12-008}, 
which may also imply a non-SM Higgs-like object.

Technicolor (TC)~\cite{Weinberg:1975gm,Farhi:1980xs} accommodates  
the electroweak symmetry breaking by techni-fermion condensation, without invoking the fundamental Higgs boson,  
just like the quark condensation in QCD, and hence gives the dynamical explanation for the origin of mass. 
The original version of TC~\cite{Weinberg:1975gm}, a naive scale-up version of QCD,  was ruled out due to the excessive 
flavor-changing neutral currents (FCNC). 
A way out of the FCNC problem was suggested under a simple assumption of the existence of 
 a large anomalous dimension for techni-fermion bilinear operator $\gamma_m$ 
without any concrete dynamics and concrete  
 value of the anomalous dimension~\cite{Holdom:1981rm}.  
It  was the walking TC (WTC)~\cite{Yamawaki:1985zg,Bando:1986bg} that exhibited a concrete dynamics based on 
a nonperturbative analysis of ladder Schwinger-Dyson (SD) equation with
nonrunning (scale invariant/conformal) gauge coupling, $\alpha(p) \equiv \alpha$, 
yielding a concrete value of the anomalous dimension, $\gamma_m = 1$ 
in the broken phase $\alpha > \alpha_c$  
where $\alpha_c$ is the 
critical coupling for the chiral symmetry breaking. 
Modern version of WTC~\cite{Lane:1991qh,Appelquist:1996dq, Miransky:1996pd} is based on the two-loop running coupling 
with the Caswell-Banks-Zaks infrared fixed point (CBZ-IRFP)~\cite{Caswell:1974gg},
 instead of the nonrunning one, in the improved ladder SD equation.

Another problem of the TC as a QCD scale-up is the electroweak constraints, so-called $S$ and $T$  parameters. 
This may also be improved in the WTC~\cite{Appelquist:1991is,Harada:2005ru}.  
Even if  WTC in isolation cannot overcome this problem, there still exist a possibility that the problem may be 
resolved in the combined dynamical system including the SM fermion mass generation such as the extended TC 
(ETC) dynamics~\cite{Dimopoulos:1979es}, 
in much the same way as the solution (``ideal fermion delocalization'')~\cite{Cacciapaglia:2004rb} in the Higgsless models which simultaneously
adjust $S$ and $T$ parameters by incorporating the SM fermion mass profile.

In the WTC the techni-fermion ($F$) acquires the mass $m_F$ dynamically 
due to the scale-invariant/conformal dynamics in the form of 
essential-singularity scaling, Miransky scaling~\cite{Miransky:1984ef}: 
\begin{equation} 
 m_F \sim \Lambda \, e^{-\pi/\sqrt{\alpha/\alpha_c -1}} 
 \, , \label{MS}
\end{equation} 
where $\Lambda$ is an ultraviolet cutoff to be identified with an ETC scale, $\Lambda \sim \Lambda_{\rm ETC}$. 
Thanks to the Miransky scaling, 
$m_F$ can be much smaller than $\Lambda$, $m_F \ll \Lambda$,  
near the criticality $\alpha \simeq \alpha_c$. 
This mass generation spontaneously breaks the scale symmetry, which 
can be characterized by the conformal phase transition~\cite{Miransky:1996pd}. 
Actually, 
once the mass $m_F$ of the techni-fermion ($F$) is dynamically generated, 
the coupling $\alpha$ starts to run slowly (``walking'') 
according to the Miransky scaling Eq.(\ref{MS})
leading to the nonperturbative beta function~\cite{Bardeen:1985sm}, 
\begin{equation} 
\beta_{\rm NP}(\alpha) 
= \frac{d \alpha}{d \log(\Lambda/m_F)}
\sim 
- \frac{1}{\left( \log \Lambda/ m_F \right)^3}
\sim -  (\alpha/\alpha_c - 1)^{3/2} 
\qquad (\alpha>\alpha_c) 
\,. \label{NP:beta}
\end{equation}  
The scale symmetry is thus spontaneously and explicitly broken by the dynamical mass generation with the scale $m_F$.

In the case of modern version of WTC~\cite{Lane:1991qh,Appelquist:1996dq, Miransky:1996pd} based on the CBZ-IRFP 
$(\alpha_*)$, the gauge coupling $\alpha$ is almost nonrunning 
$\alpha(p) \sim {\rm cons.} \simeq \alpha_*$ for $m_F < p< \Lambda_{\rm TC}$ 
where $\Lambda_{\rm TC}$ is an intrinsic scale analogous to $\Lambda_{\rm QCD}$ 
and plays a role of the ultraviolet cutoff $\Lambda$: $\Lambda \sim \Lambda_{\rm TC} \sim \Lambda_{\rm ETC} 
(>10^3 \, {\rm TeV})$.    
The existence of the intrinsic scale $\Lambda_{\rm TC}$ 
breaks the scale symmetry already at two-loop perturbative level for the ultraviolet region $p>\Lambda_{\rm TC}$, 
where the coupling runs in the same way as in QCD. 
However, this perturbative scale-symmetry-breaking scale $\Lambda_{\rm TC}$ is 
irrelevant to the dynamical mass $m_F$ which can be much smaller than $\Lambda_{\rm TC}$.  
The very reason for the nonperturbative scale anomaly thus comes only from the dynamical fermion mass generation 
along with the infrared scale $m_F$.

The spontaneous breaking of such an approximate scale symmetry implies 
existence of a light composite scalar, techni-dilaton (TD)~\cite{Yamawaki:1985zg,Bando:1986bg}, 
arising as the pseudo Nambu-Goldstone boson (pNGB) for the scale symmetry.  
Since the TD field ($\phi$) couples to the trace of energy momentum tensor $\theta_\mu^\mu$ which is chiral invariant, 
the composite TD is formed by techni-fermion and anti-techni-fermion bound state $(\bar{F}F)$ in a chiral- and flavor-singlet manner: 
\begin{equation} 
  \bar{F}F \approx \langle \bar{F}F \rangle e^{(3-\gamma_m) \phi/F_{\phi}}\cdot  U     
\,,  \label{FF-phi}
\end{equation}  
where $F_\phi$ is the TD decay constant related to the spontaneous breaking of the scale symmetry; 
$(3-\gamma_m) \simeq 2$ denotes scale dimension of $\bar{F}F$; $U$ is the usual chiral field parametrized by 
the NGB fields associated with the spontaneously broken chiral symmetry as $U=e^{2i \pi/F_\pi}$ with the techni-pion 
decay constant $F_\pi$.

The TD gets massive due to the nonperturbative scale anomaly as mentioned above, 
generated from the nonperturbative renormalization of the TC gauge coupling $\alpha$ 
as in Eq.(\ref{NP:beta}) 
associated with the techni-fermion mass 
generation via Miransky scaling. 
The TD mass $M_\phi$ and decay constant $F_\phi$ may then be related 
to $\theta_\mu^\mu$, trace of the energy momentum tensor, through the partially 
conserved dilatation current (PCDC) for the trace anomaly: 
\begin{equation} 
 F_\phi^2 M_\phi^2 = - 4 \langle \theta_\mu^\mu \rangle 
 \,, \qquad 
 \theta_\mu^\mu 
 = \frac{\beta_{\rm NP}(\alpha)}{4 \alpha} G_{\mu\nu}^2 
\,, \label{PCDC} 
\end{equation}
where $G_{\mu\nu}$ stands for the techni-gluon field strength. 
Here $\langle \theta_\mu^\mu \rangle = 4 {\cal E}_{\rm vac}$, with ${\cal E}_{\rm vac}$ being the vacuum energy density,  
and ${\cal E}_{\rm vac}$ only includes contributions from the nonperturbative scale anomaly, defined by subtracting contributions $\langle \theta_\mu^\mu \rangle_{\rm perturbation}$ of ${\cal O}(\Lambda_{\rm TC}^4)$ 
from the perturbative running of the gauge coupling $\alpha$, such as 
$\langle \theta_\mu^\mu  \rangle  - \langle \theta_\mu^\mu \rangle_{\rm perturbation}$, 
which is saturated by the techni-gluon condensation induced by the techni-fermion condensation.   
Hence the PCDC relation Eq.(\ref{PCDC}) generically scales like  
\begin{equation} 
  F_\phi^2 M_\phi^2 
  = - 16 {\cal E}_{\rm vac}
  \sim 
16 \left( \frac{d_F N_{\rm TF}}{\pi^4} \right) m_F^4 
  \,, \label{PCDC:gene}
\end{equation} 
where $d_F$ is dimension of techni-fermion representation for $SU(N_{\rm TC})$ TC gauge group (say, $d_F=N_{\rm TC}$ for 
fundamental representation) and $N_{\rm TF}$ denotes the number of techni-fermions.  
Eq.(\ref{PCDC:gene}) indeed implies that TD can  be lighter than other hadrons having masses of ${\cal O}(m_F)$ 
when the scale of $F_\phi$ is of ${\cal O}(m_F)$ or higher. 
However, the TD mass cannot parametrically be small since 
the scale symmetry is actually broken explicitly 
for  the very reason of the spontaneous breaking itself, 
namely the dynamical generation of the techni-fermion mass $m_F$, which 
is  responsible for the nonperturbative running of $\alpha$ (nonperturbative scale anomaly Eq.(\ref{PCDC})) as mentioned above. 
In fact, straightforward nonperturbative calculations of ${\cal E}_{\rm vac}$ based on the ladder SD 
analysis~\cite{Miransky:1989qc,Hashimoto:2010nw} support 
the scaling Eq.(\ref{PCDC:gene}) and show that $(F_\phi/m_F) \cdot (M_\phi/m_F) =$ finite even at the 
criticality limit $\alpha \to \alpha_c$ $(m_F/\Lambda_{\rm TC} \to 0 )$ where $\beta_{\rm NP}(\alpha) \to 0$. 
Thus the TD cannot be massless  
unless it is decoupled by $F_\phi \to \infty$~\footnote{Such a ``decoupled TD'' 
scenario~\cite{Haba:2010hu,Hashimoto:2010nw} with the Yukawa coupling  
$\sim m_F/ F_{\phi} \to 0 $ as $m_F/\Lambda_{\rm TC} \to 0$ 
might be relevant to dark matter~\cite{Hashimoto:2010nw,Choi:2011fy}.}.

In fact, $M_{\phi} \simeq 500 - 600 \, {\rm GeV}$ for the typical one-family model was suggested~\cite{Yamawaki:2007zz}, 
based on various explicit calculations~\cite{Shuto:1989te}. 
(This is also consistent with the recent holographic estimate~\cite{Haba:2010hu} and others~\cite{Kutasov:2011fr}.) 
However, their results are not very conclusive due to the respective uncertainties in those computations. 
Although such a composite scalar 
was identified as a chiral non-singlet state, just like the chiral partner of pion,  
in contrast to the correct identification of TD as in Eq.(\ref{FF-phi}), 
the TD mass would be more involved than  
such an estimated scalar mass. 
It may therefore be reasonable to deal with the TD mass as a free parameter at present.

Actually, we have recently explored the TD LHC signatures taking 
the mass as a free parameter in the range  
110 GeV -- 1000 GeV, which is 
in the reach of LHC experiments~\cite{Matsuzaki:2011ie,Matsuzaki:2012gd}. 
We addressed how the signatures look different from those of the SM Higgs for 
by explicitly calculating the TD LHC production cross sections at 7 TeV 
times branching ratios normalized to the corresponding quantities for the SM Higgs. 
 Particularly in Ref.~\cite{Matsuzaki:2012gd} 
the currently observed excess at 125 GeV 
in the diphoton channel can be explained by TD.

In this paper, we first refine the previous calculations of Refs.~\cite{Matsuzaki:2011ie,Matsuzaki:2012gd} 
and~\cite{Hashimoto:2011cw}, 
based on a new method to derive all the 
TD couplings to the SM particles and techni-fermions, solely from 
the Ward-Takahashi identities for the dilatation current coupled to TD.  
We show that all the couplings to the SM particles are induced from techni-fermion loops: 
The Yukawa couplings to the SM fermions arise due to  
ETC-induced four-fermion interactions reflecting the ultraviolet feature of WTC 
characterized by the anomalous dimension $\gamma_m \simeq 1 $.  
The couplings to the SM gauge bosons, on the other hand, are determined 
 by the infrared features fixed solely by the low-energy theorem.

We also refine the low-energy effective Lagrangian for TD in a way consistent 
with the Ward-Takahashi identities mentioned above. 
The Lagrangian is 
based on the nonlinear realization of both the scale and chiral symmetries, where 
the scale invariance is ensured by including a spurion field which reflects 
the explicit breaking induced from the dynamical generation of techni-fermion mass itself. 
We then discuss the stability on the TD mass against quadratically divergent corrections arising from an 
effective theory below the scale $m_F$ which would be the only possible source for a sizable scale symmetry breaking relevant 
to the TD mass.

To be concrete, we take typical models of WTC such as one-doublet model (1DM) and 
one-family model (1FM) to discuss the TD LHC signatures in comparison with those of the SM Higgs  
by changing the TD mass as a free parameter concentrated on a light mass region 
110 GeV - 600 GeV.   
 It turns out that the TD in the 1DM is invisible   
due to the highly suppressed couplings to the SM particles. 
It is shown that the light TD signal in the 1FM can be found as a large excess  
relative to the SM Higgs 
at around 125 GeV   
only in the diphoton channel.

This paper is organized as follows: 
In Sec.~\ref{TDcouplings} we present a new derivation of all the TD couplings 
directly from the Ward-Takahashi identities. 
In Sec.~\ref{nonlinear} 
we refine the low-energy effective Lagrangian for TD in a way consistent with 
the Ward-Takahashi identities. 
In Sec.~\ref{TDLHC} we discuss the TD LHC signatures for the TD mass range 110 GeV $\le M_\phi \le$ 600 GeV, 
taking the 1FM 
to make definite benchmarks.  
Sec.~\ref{summary} is devoted to summary of this paper. 
The formulas for the TD partial decay widths 
are presented in Appendix~\ref{widths}.

\section{TD couplings} 
\label{TDcouplings}

In this section, we shall derive formulas for the TD couplings to the techni-fermions and SM particles 
through the Ward-Takahashi identities for the dilatation current coupled to TD.  
It is shown that all the couplings to the SM particles are induced from techni-fermion loops: 
The Yukawa couplings to the SM fermions arise due to  
ETC-induced four-fermion interactions reflecting the ultraviolet feature of WTC 
characterized by the anomalous dimension $\gamma_m \simeq 1 $ (Sec.~\ref{CSMF}).  
The couplings to the SM gauge bosons, on the other hand, are determined 
 by the infrared features fixed solely by the low-energy theorem (Sec.~\ref{CSMG}).

\subsection{Coupling to the techni-fermions} 
\label{CTF}

We start with a low-energy theorem related to 
the Ward-Takahashi identity for a techni-fermion two-point function coupled to the 
dilatation current $D_\mu=\theta_{\mu\nu} x^\nu$:  
\begin{eqnarray}
 \lim_{q_\mu \to 0}  \,  
\int d^4 y \, e^{i qy} 
 \langle 0 |T \partial^\mu D_\mu(y) F(x) \bar{F}(0)  |0 \rangle 
&=&  
 i \delta_D \langle 0 |T F(x)\bar{F}(0)  |0 \rangle 
 \nonumber \\ 
&=& i \left( 2 d_F 
   +  x^\nu \partial_\nu  \right) \langle 0| T F(x) \bar{F}(0)  |0 \rangle 
   \,, \label{WT:FFbar}
\end{eqnarray}
where we used $[i Q_D, F(x)]=\delta_D F(x)=(d_F + x^\nu \partial_\nu) F(x)$ 
with the dilatation charge $Q_D=\int d^3 x D_0(x)$, 
in which $d_F=3/2$ denotes the scale dimension of techni-fermion field $F$. 
Assuming TD-pole dominance in the left-hand side, we rewrite Eq.(\ref{WT:FFbar}) as 
\begin{equation} 
F_\phi \cdot \langle \phi(q=0) |T  F(x) \bar{F}(0)  |0 \rangle 
= 
 \delta_D \langle 0 |T F(x)\bar{F}(0)  |0 \rangle 
\,, 
\end{equation}
where use has been made of the definition of the TD decay constant $F_\phi$: 
\begin{equation} 
 \langle 0| D_\mu (x) | \phi(q) \rangle 
 = - i F_\phi q_\mu e^{-iq x} 
 \,, \label{Fphi:def}
\end{equation}
in which $D_\mu$ stands for the dilatation current constructed only from the TC sector fields.  
 Taking Fourier transform (F.T.)  of both sides with momentum $p$, we find 
\begin{eqnarray} 
 {\rm F.T.} \langle \phi(q=0) |T  F(x) \bar{F}(0)  |0 \rangle  
 &=& - \frac{1}{F_\phi} 
\delta_D S_F(p) 
=
\frac{1}{F_\phi} \left[
   S_F(p) +  p_\mu \frac{\partial}{\partial p_\mu} S_F(p) 
\right]
\nonumber \\ 
&=& 
\frac{1}{F_\phi} S_F(p) \cdot \left[  
 \delta_D S_F^{-1}(p) 
\right] 
\cdot S_F(p) 
\,, \label{WT}
\end{eqnarray}  
where 
\begin{equation} 
 \delta_D S_F^{-1} (p) 
 = 
\left( 1 -  p_\mu \frac{\partial}{\partial p_\mu} \right) S_F^{-1}(p) 
 \,, 
\end{equation}
with $S_F(p)$ being the (full) propagator of techni-fermion defined as 
$S_F(p) = {\rm F.T.} \langle 0 |T  F(x) \bar{F}(0)  |0 \rangle  \equiv \int d^4x e^{ipx} 
 \langle 0 |T  F(x) \bar{F}(0)  |0 \rangle$. 
 We shall define the amputated Yukawa vertex function $\chi_{\phi FF}(p, q)$: 
\begin{equation} 
  \chi_{\phi FF}(p, q)
\equiv 
S_F^{-1}(p) \cdot \left( {\rm F.T.} \langle \phi(q) |T  F(y) \bar{F}(0)  |0 \rangle  \right) \cdot S_F^{-1}(p+q)
\,. \label{def:chi}
\end{equation} 
Eq.(\ref{WT}) then reads  
\begin{equation} 
 \chi_{\phi FF}(p, q=0) 
 = \frac{1}{F_\phi} \delta_D S_F^{-1}(p) 
 = \frac{1}{F_\phi} \left( 1 -  p_\mu \frac{\partial}{\partial p_\mu} \right) S_F^{-1}(p) 
 \,. \label{chi:FF} 
\end{equation}

As done in the original literature~\cite{Bando:1986bg}, 
one may also derive a Ward-Takahashi identity for a local composite operator $\bar{F}F(0)$ 
having the scale dimension $(3-\gamma_m)$: 
\begin{eqnarray} 
 \lim_{q_\mu \to 0} \, \int d^4x e^{iqx} \langle 0 | T \partial_\mu D^\mu(x) \bar{F} F(0)  \rangle
 &=&  i \delta_D \langle \bar{F}F \rangle   
 =
i (3-\gamma_m) \langle \bar{F}F \rangle 
 \nonumber \\ 
 \to \qquad  
 \langle 0| \bar{F}F(0) | \phi(q=0) \rangle 
 &=& 
 - \frac{1}{F_\phi} \delta_D  \langle \bar{F}F \rangle 
 = 
 - \frac{(3-\gamma_m)}{F_\phi} \langle \bar{F}F \rangle
 \,,  \label{WT:FF}
\end{eqnarray}
which implies an operator relation between $\bar{F}F$, $\phi$ and $U=e^{2i\pi/F_\pi}$ as given in Eq.(\ref{FF-phi}), 
\begin{equation} 
  \bar{F}F \approx \langle \bar{F}F \rangle e^{-(3-\gamma_m)\phi/F_\phi} \cdot U
\,, \label{FF-phi:2}
\end{equation}
with the normalization of $\phi$-state as $\langle 0 | \phi | \phi \rangle = 1$ and $\langle 0| U |0 \rangle=1$. 

The TD decay constant $F_\phi$ can be related to the TD mass through the PCDC relation 
based on the Ward-Takahashi identity associated with the trace of energy-momentum tensor $\theta_\mu^\mu=\partial_\mu D^\mu$, similarly to Eq.(\ref{WT:FFbar}): 
\begin{eqnarray}
 \lim_{q_\mu \to 0} \, \int d^4 x e^{ i qx} 
 \langle 0 |T \partial_\mu D^\mu(x) \theta_\nu^\nu(0)  |0 \rangle 
 = i \delta_D  \langle 0| \theta_\mu^\mu(0)  |0 \rangle 
 = 
i d_\theta \, \langle 0| \theta_\mu^\mu(0)  |0 \rangle 
   \,, \label{WT:theta}
\end{eqnarray}
where $d_\theta (= 4)$ is the scale dimension of $\theta_\mu^\mu$. 
 The TD-pole dominance and use of Eq.(\ref{Fphi:def}) thus lead to 
 the PCDC relation~\footnote{
 Note that if one wrote an operator relation like ``$\partial_\mu D^\mu(x) = F_\phi M_\phi^2 \phi(x)$",  
$\phi(x)$ would mean merely a generic scalar density as 
an interpolating field of TD as in the case of PCAC (partially conserved axialvector current): 
The PCDC relation should {\it not} be understood as 
an operator relation. }, 
\begin{equation} 
  F_\phi^2 M_\phi^2 = - d_\theta \langle \theta_\mu^\mu \rangle
  \,, \label{PCDC:relation}
\end{equation} 
 which will be used for the phenomenological studies of TD given in the later sections.

\subsection{Coupling to the SM fermions} 
  \label{CSMF}

\begin{figure}

\begin{center} 
\includegraphics[scale=0.7]{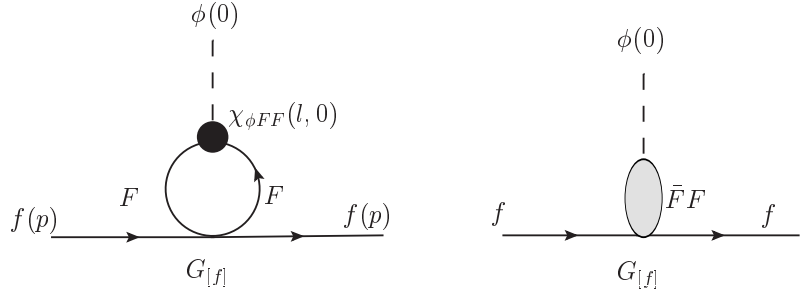}
\end{center}
\caption{ 
Left panel: The Feynman graph corresponding to the amplitude Eq.(\ref{Mff}) which generates 
the Yukawa vertex for the SM $f$-fermion. 
 Right panel: The graphical interpretation for derivation of Eq.(\ref{yukawa:f:2}) through ETC-induced four-fermion interaction in Eq.(\ref{4fermi})
with the coupling strength $G_{[f]}$ coupled to the composite $\bar{F}F$ operator. }  
\label{yukawa-ff}
\end{figure}

Since the dilatation current $D_\mu$ in Eq.(\ref{Fphi:def}) consists only of the TC sector fields,  
all the SM fermion fields do not transform under the scale symmetry, $\delta_D f(x) =0$.  
Accordingly, they do not directly couple to TD: 
\begin{equation} 
 \langle f (p) | \theta_\mu^\mu(0) | f(p) \rangle 
 = 0 
 \,. \label{yukawa:p}
\end{equation}
Their couplings are thus generated only through an ETC contribution 
communicating the TC sector to the SM fermion sector. 
They can be described in a low-energy effective Lagrangian as  
\begin{equation} 
  {\cal L}_{\rm ETC}^{\rm eff} = G_{[f]} \bar{F} F \bar{f} f 
  \,, \label{4fermi}
\end{equation}
which gives the $f$-fermion mass through the techni-fermion condensation: 
\begin{equation} 
m_f= - G_{[f]} \langle \bar{F}F \rangle
\,. \label{mf}
\end{equation}  
  The interaction term in Eq.(\ref{4fermi}) 
together with the Yukawa vertex function Eq.(\ref{chi:FF}) 
gives rise to the following matrix element (See also the left panel of Fig.~\ref{yukawa-ff}): 
 \begin{eqnarray} 
  i {\cal M}(\phi(0), f(p), \bar{f}(p)) 
  &=&   - \frac{ i G_{[f]}}{F_\phi} \int \frac{d^4 l}{(2 \pi)^4} {\rm Tr} 
[ S_F(l) \cdot 
\delta_D S_F^{-1}(l) 
\cdot  
 S_F(l) ] \cdot \bar{u}_f(p) u_f(p) 
\nonumber \\ 
&=& 
  \frac{ i G_{[f]}}{F_\phi} \cdot \delta_D 
\int \frac{d^4 l}{(2 \pi)^4} {\rm Tr} [ 
S_F(l) ] 
\cdot \bar{u}_f(p) u_f(p) 
\,, \label{Mff}
 \end{eqnarray}
 where $u_f(p)$ denote the wavefunction of the SM-$f$ fermion field. 
Noting that 
\begin{eqnarray} 
\langle  \bar{F}F \rangle
= - \int \frac{d^4 l}{(2\pi)^4} {\rm Tr}[S_F(l)] 
 \,, 
\end{eqnarray}
and Eq.(\ref{mf}), we find   
\begin{equation}
  i {\cal M}(\phi(0), f(p), \bar{f}(p))
  = 
  - i \frac{G_{[f]}}{F_\phi} \delta_D \langle \bar{F}F \rangle
  \cdot \bar{u}_f(p) u_f(p)  
=  i \frac{(3-\gamma_m) m_f}{F_\phi}  \bar{u}_f(p) u_f(p) 
  \,, 
\end{equation}
which gives the Yukawa coupling to the SM $f$-fermion in a view of an effective Lagrangian, 
\begin{equation} 
{\cal L}_{\phi ff} 
= g_{\phi ff} \, \phi \bar{f}f 
\,, \qquad 
  g_{\phi ff} = \frac{(3-\gamma_m) m_f}{F_\phi} 
  \,. \label{yukawa:f:1}
\end{equation}

As was done in Ref.~\cite{Bando:1986bg}, 
one can reach the same formula as Eq.(\ref{yukawa:f:1}) by considering 
the composite $\bar{F}F$ operator insertion, as illustrated in the right panel of Fig.~\ref{yukawa-ff}: 
Using the operator relation between $\bar{F}F$ and $\phi$ given in Eq.(\ref{FF-phi:2}) consistently with 
the Ward-Takahashi identity Eq.(\ref{WT:FF}), 
Eq.(\ref{4fermi}) reads    
\begin{equation} 
 G_{[f]} \bar{F}F \bar{f}f \approx 
  - m_f \bar{f}f + \frac{(3-\gamma_m) m_f}{F_\phi} \, \phi \bar{f}f + \cdots. 
\label{yukawa:f:2}  
\end{equation}

\subsection{Couplings to the SM gauge bosons}  
\label{CSMG}

The TD couplings to the SM gauge bosons are also 
generated only through the techni-fermion loops. 
We shall first consider a low-energy theorem associated with 
the Ward-Takahashi identity for a techni-fermion vector/axialvector current $J_\mu$  
coupled to the trace of energy-momentum tensor $\theta_\mu^\mu=\partial^\mu D_\mu$: 
\begin{eqnarray} 
 \lim_{q_\rho \to 0} \int d^4z \, e^{iqz} \, \langle 0|T \partial_\rho D^\rho(z) J_\mu(x) J_\nu(0)  |0 \rangle 
  &=& \lim_{q_\rho \to 0} \left( -  i q_\rho \int d^4z \, e^{iqz} \, \langle 0| T D^\rho(z) J_\mu(x) J_\nu(0) |0 \rangle  \right)
\nonumber \\ 
&& 
+ i \delta_D  \langle 0| T J_\mu(x) J_\nu(0) |0\rangle
\,, 
\end{eqnarray}
where $\delta_D \langle 0| T J_\mu(x) J_\nu(0) |0\rangle
=(2 d_J + x^\rho \partial_\rho^x) \langle 0| T J_\mu(x) J_\nu(0) |0\rangle 
$ with 
$d_J(=3)$ being the scale dimension of the current $J_\mu$. 
Here all suffixes regarding the current $J_\mu$ such as the TC and SM charges  
have been suppressed for simplicity. 
In the first line of the right hand side, the scale/dilatation anomaly induced 
by techni-fermion loops has been incorporated properly. 
 Taking the Fourier transform of both sides and extracting the dilaton pole from the left hand side 
by using Eq.(\ref{Fphi:def}), we find 
\begin{equation} 
  {\rm F.T.} \langle \phi(0) | T J_\mu(x) J_\nu(0) |0 \rangle 
  = \frac{1}{F_\phi} \left\{ 
  \lim_{q_\rho \to 0} q_\rho \Gamma_{\mu\nu}^\rho(p, q-p;q) 
  - 
\delta_D \Pi_{\mu\nu}(p) 
\right\}  
\,, \label{WT:JJ}
\end{equation}
  where 
\begin{eqnarray} 
  \Gamma_{\mu\nu}^\rho(p, q-p;q) 
  &=& \int d^4z \, dx^4  \, e^{iqz - i px} \, \langle 0| T D^\rho(z) J_\mu(x) J_\nu(0) |0 \rangle 
  \,, \nonumber \\ 
  \Pi_{\mu\nu}(p) 
&=&  
\int dx^2 x \, e^{- i px} \, \langle 0| T J_\mu(x) J_\nu(0) |0 \rangle 
\,, \nonumber \\
\delta_D \Pi_{\mu\nu}(p) 
&=& \left( (2d_J -4)  - p_\rho \frac{\partial}{\partial p_\rho} \right) \Pi_{\mu\nu}(p) 
\,. 
\end{eqnarray}

The anomaly-free term, the second term of the right hand side in Eq.(\ref{WT:JJ}), may further be rewritten into the form 
\begin{eqnarray} 
  {\rm F.T.} \langle \phi(0) | T J_\mu(x) J_\nu(0) |0 \rangle \Bigg|_{\rm anomaly-free}
&=& - \frac{1}{F_\phi} \delta_D \Pi_{\mu\nu}(p) 
\nonumber \\ 
&=&
- \frac{2 i}{F_\phi} \left(g_{\mu\nu} - \frac{p_\mu p_\nu}{p^2} \right) 
\left[ 1 - p^2 \frac{\partial}{\partial p^2} \right] \Pi(p^2) 
\,, 
\end{eqnarray} 
where we defined the current correlator $\Pi(p^2)$ as 
$\Pi_{\mu\nu}(p) = i (g_{\mu\nu} - \frac{p_\mu p_\nu}{p^2}) \Pi(p^2)$. 
Consider the $SU(2)_W$ current $J_L^{\mu a}= \bar{F}_L \gamma^\mu \frac{\sigma^a}{2} F_L$ 
with $\sigma^a$ $(a=1,2,3)$ being the Pauli matrices. 
We may further expand the current correlator $\Pi_{LL}(p^2)$ around $p^2=0$ to find  
\begin{equation} 
  {\rm F.T.} \langle \phi(0) | T J_L^{\mu a}(x) J_L^{\nu b}(0) |0 \rangle \Bigg|_{\rm anomaly-free}
  = 
  - \delta^{ab} \frac{2 i}{F_\phi} \left(g_{\mu\nu} - \frac{p_\mu p_\nu}{p^2} \right) 
\left[ \Pi_{LL}(0) + {\cal O}(p^4 \Pi^{\prime\prime}(0)) \right]   
\,.  
\end{equation} 
Since the $SU(2)_W$ current is spontaneously broken to be coupled to the associated NGBs, 
one should find the decay constant $F_\pi$ in $\Pi_{LL}(0)$: 
\begin{equation} 
  \Pi_{LL}(0) = N_D \frac{F_\pi^2}{4} = \frac{v_{\rm EW}^2}{4}
  \,, 
\end{equation}
  where $N_D$ is the number of the $SU(2)_W$ doublets formed by the techni-fermions. 
Supplying the $SU(2)_W$ gauge coupling $g_W$ to the $SU(2)_L$ current coupled to the SM weak boson 
and identifying the resultant amplitude as the $\phi$-$W^a$-$W^b$ vertex function such that 
\begin{equation} 
i g_W^2 \,{\rm F.T.} \langle \phi(0) | T J_L^{\mu a}(x) J_L^{\nu b}(0) |0 \rangle \Bigg|_{\rm anomaly-free}
\equiv  \delta^{ab}  \left(g_{\mu\nu} - \frac{p_\mu p_\nu}{p^2} \right) g_{\phi WW}(p^2)
\,, 
\end{equation} 
we thus arrive at 
\begin{equation} 
 g_{\phi WW}(0) = \frac{2 m_W^2}{F_\phi} 
 \,. 
\end{equation} 
Note that 
this result reflects the low-energy behavior of 
the TD Yukawa vertex corresponding to ``$3-\gamma_m =1$" 
in comparison with Refs.~\cite{Matsuzaki:2011ie,Matsuzaki:2012gd}~\footnote{
 The TD Yukawa vertex in Eq.(\ref{chi:FF}) 
 behaves as $\chi_{\phi FF}(p) = \delta S_F^{-1}(p)/F_\phi \sim \Sigma(p^2)/F_\phi \sim p^{\gamma_m -2}$ 
 in the asymptotic region $p^2 \ge m_F^2$, where $\Sigma(p^2)$ denotes 
 the mass function. 
Since 
the ultraviolet region is highly suppressed as $ 
I \sim   \int d^4 p \, \frac{\Sigma(p^2)^2}{p^4} \sim \int d^4 p\, p^{2\gamma_m -8}$ 
in the relevant loop integral $g_{\mu\nu} \cdot I$ for the TD-$W$-$W$ vertex, 
it is dominated by the infrared region where the Yukawa vertex is almost constant 
corresponding to $\gamma_m =2$ ($3-\gamma_m=1$). It is also the case 
 for the scale anomaly term in Eq.(\ref{scale-anomaly}).  
\label{infra}   
}. 
In terms of an effective Lagrangian the coupling can be viewed as 
\begin{equation} 
  {\cal L}_{\phi WW} 
= \frac{2 m_W^2}{F_\phi}  \phi W_\mu^a W^{\mu a} 
\,. \label{coupling:WW}
\end{equation}

Similarly, one can easily derive the coupling formula for the $U(1)_Y$ gauge boson 
and apply the standard weak mixing to the weak gauge bosons to get the coupling to the $Z$ boson: 
\begin{eqnarray} 
 g_{\phi ZZ} 
= \frac{2  m_Z^2}{F_\phi} 
\,, \qquad 
  {\cal L}_{\phi ZZ} 
= 
\frac{1}{2}
g_{\phi ZZ} \, \phi Z_\mu Z^{\mu} 
 \,. \label{coupling:ZZ}
\end{eqnarray}
 Eqs.(\ref{coupling:WW}) and (\ref{coupling:ZZ}) thus imply 
that the TD couplings to the weak gauge bosons take essentially the same form 
as those of the SM Higgs, except for the overall scale set to $F_\phi$, instead of $v_{\rm EW}$.

\begin{figure}

\begin{center} 
\includegraphics[scale=0.35]{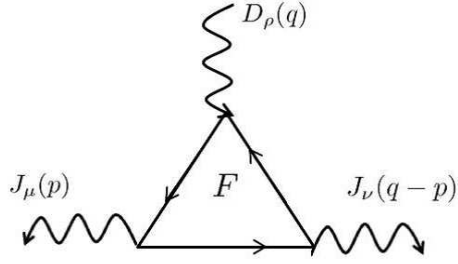}
\end{center}
\caption{ 
A triangle-loop graph yielding the scale anomaly induced by the techni-fermion loop. }  
\label{triangle}
\end{figure}

 On the other hand, 
the couplings to the other gauge bosons such as gluon and photon 
do not arise from the anomaly-free term at leading order of derivative expansion, 
since they couple to unbroken currents where $\Pi(0)=0$. 
Those couplings actually come from the anomaly term,  
the first term of the right hand side in Eq.(\ref{WT:JJ}).

The anomaly term 
can be calculated by straightforwardly 
evaluating the triangle diagram as depicted in Fig.~\ref{triangle}:  
\begin{equation}
 \lim_{q_\rho \to 0} \, 
q_\rho \Gamma_{\mu\nu}^\rho(p,q-p;q) 
  = i \frac{2 \beta_F(g)}{ g^3} (p^2 g_{\mu\nu} - p_\mu p_\nu)
  \,, \label{scale-anomaly}
\end{equation}
where $g$ stands for some gauge coupling associated with a gauge boson coupled to the current $J_\mu$ 
and $\beta_F(g)$ the corresponding beta function including only the techni-fermion contributions.  
 In terms of an effective Lagrangian, the vertex function can be viewed as 
\begin{equation} 
  {\cal L}_{\phi VV} 
= \frac{\beta_F(g)}{2g^3} \, \frac{\phi}{F_\phi} V_{\mu\nu}^2  
\,, 
\end{equation} 
 where $V_{\mu\nu}$ is a field strength for an SM gauge field $V_\mu$. 
 For instance, the TD couplings to $\gamma\gamma$ and $gg$ read 
\begin{eqnarray} 
 {\cal L}_{\phi \gamma\gamma, gg} 
 &=& \frac{\phi}{F_\phi}  \left[ 
\frac{\beta_F(e)}{2 e^3} F_{\mu\nu}^2 
 + 
 \frac{\beta_F(g_s)}{2 g_s^3} G_{\mu\nu}^2 
\right]
\,, \label{TD:dip-dig}
\end{eqnarray}
with the electromagnetic and QCD couplings $e$ and $g$ and their field strengths. 
In addition to the techni-fermion loop contributions, the $W$ boson and the SM fermion loop corrections 
 should be incorporated in fully evaluating the vertex functions. 
 For the full expressions, see Appendix~\ref{widths}.

Note that our results obtained in this paper are  direct consequences of the Ward-Takahashi identities 
without referring to the explicit form of the TD Yukawa coupling to techni-fermions.

\section{Phenomenological Lagrangian for TD}
\label{nonlinear}

In this section, 
we shall introduce an effective Lagrangian to reproduce the results 
obtained in Sec.~\ref{TDcouplings} on the TD couplings to the SM particles 
when the low-energy limit $p \ll m_F$ is taken (Sec.~\ref{nonlinearLag}).  
We also discuss the stability on the TD mass against quadratically divergent corrections arising from an 
effective theory below the scale $m_F$ which would be the only possible source for a sizable scale symmetry breaking relevant 
to the TD mass (Sec.~\ref{naturalness}).

\subsection{Nonlinear realization} 
\label{nonlinearLag}

We begin by introducing the TD and techni-pion fields, $\phi$ and $\pi$, nonlinearly transforming under 
the scale and chiral $SU(N_{\rm TF})_L \times SU(N_{\rm TF})_R$ symmetries, respectively. 
The techni-pion fields $\pi$ are embedded into the usual chiral nonlinear base $U$ parametrized 
as $U=e^{2i \pi/F_\pi}$ where $\pi=\pi^A X^A$ $(A=1,\cdots ,N_{\rm TF}^2-1)$ with $X^A$ being 
generators of $SU(N_{\rm TF})$ such that ${\rm Tr}[X^AX^B]=\delta^{AB}/2$, 
and $F_\pi$ denotes the decay constant associated with the spontaneous 
breaking of the chiral symmetry. 
The chiral nonlinear base $U$ then transforms under the chiral symmetry as $U \to g_L U g_R^\dag$, 
while under the scale symmetry $\delta U= x^\nu \partial_\nu U$ and so does $\pi$. 
The TD field $\phi$ is, on the other hand, related to 
a field $\chi$ transforming linearly under 
the scale transformation, such that 
\begin{equation} 
\chi=e^{\phi/F_\phi} 
\, , \qquad 
\delta \chi = (1 + x^\nu \partial_\nu) \chi 
\,, 
\end{equation}  
while $\phi$ does nonlinearly as 
\begin{equation} 
\delta \phi= F_\phi + x^\nu \partial_\nu \phi 
\, .  
\end{equation}

The kinetic terms for the techni-pions and TD thus  
take the scale-invariant form    
\begin{equation} 
  {\cal L}_{\rm inv} 
= \frac{F_\pi^2}{4} \chi^2 {\rm Tr}[{\cal D}_\mu U^\dag {\cal D}^\mu U] 
  + 
\frac{F_{\phi}^2}{2} \partial_\mu \chi \partial^\mu \chi 
  \,, \label{L:inv}
\end{equation}
where ${\cal D}_\mu U$ denotes the covariant derivative acting on $U$ gauged 
only under the SM $SU(3)_c \times SU(2)_W \times U(1)_Y$ 
gauge symmetries.

The scale symmetry is actually broken explicitly as well as spontaneously  by 
dynamical techni-fermion mass generation, which   
 has to be respected also in the nonlinear realization~\cite{Matsuzaki:2011ie}. 
  In order to incorporate these effects into the scale-invariant Lagrangian, 
we introduce a spurion field $S$ having 
the scale dimension $1$ coupled to the SM fermions, $gg$ and $\gamma\gamma$ 
in such a way that 
\begin{eqnarray} 
{\cal L}_{S} 
&=& 
- m_f \left( \left( \frac{\chi}{S} \right)^{2-\gamma_m} \cdot \chi \right) \bar{f} f 
\nonumber \\ 
&& 
+  \log \left( \frac{\chi}{S} \right) \left\{ 
\frac{\beta_F(g_s)}{2g_s} G_{\mu\nu}^2 
+ 
\frac{\beta_F(e)}{2e} F_{\mu\nu}^2 
\right\} 
+ \cdots 
\,, \label{Lag:S}
\end{eqnarray}
where $G_{\mu\nu}$ and $F_{\mu\nu}$ respectively denote the field strengths for gluon and photon fields, and 
$\beta_F$s are  the beta functions only including the techni-fermion 
loop contributions.  The ellipses in the second line would include techni-pion mass terms 
coming from ETC contributions explicitly breaking the scale symmetry. 
 However, the scale of techni-pion masses turns out to actually be above the cutoff scale~\cite{Jia:2012kd} 
set by $m_F$ which is estimated to be $\simeq 320 \, {\rm GeV} \sqrt{\frac{4}{N_D} \frac{3}{N_{\rm TC}}}$ (See Eq.(\ref{vals})). 
Hence we have not written such terms in Eq.(\ref{Lag:S}).

In addition, the TD potential term $V_\chi$ including nonderivative couplings should be incorporated so as to 
reproduce the desired nonperturbative scale anomaly Eq.(\ref{PCDC})~\cite{Schechter:1980ak}: 
\begin{equation} 
  V_\chi = \frac{F_{\phi}^2 M_{\phi}^2}{4} \chi^4 \left ( \log \chi - \frac{1}{4} \right)
  \,, \label{V:chi}
\end{equation}
One can easily check that the scale transformation of $V_\chi$ certainly yields the PCDC 
relation Eq.(\ref{PCDC}),  
\begin{equation} 
  \langle \theta_\mu^\mu \rangle 
= - \delta_D V_\chi \Bigg|_{\rm vacuum} = - \frac{F_{\phi}^2 M_{\phi}^2}{4} \langle \chi^4 \rangle \Bigg|_{\rm vacuum} = 
- \frac{F_{\phi}^2 M_{\phi}^2}{4}
\,. \label{NPanomaly}
\end{equation} 
Note that although an operator relation of the PCDC 
would be violated if one 
put $S(x)=1$ from the beginning in our Lagrangian, 
the PCDC relation should not be understood as an operator relation 
as we mentioned below Eq.(\ref{PCDC:relation}). 
Actually, 
the operator form of the PCDC relation is obtained by keeping 
  $S(x)$ as an operator.  
Matrix elements involving TD should be calculated keeping $S(x)$ 
as an operator and then putting $S(x)=1$ after all calculations, 
as in the case of other spurion methods.

\subsection{The size of radiative corrections to the TD mass} 
\label{naturalness}

Before closing this section, 
we shall briefly remark on  stability of the light TD mass against radiative corrections.  
As a pNGB of scale invariance the quadratic divergence is suppressed by the 
scale invariance for the walking regime $m_F < \mu <\Lambda_{\rm TC}(\sim \Lambda_{\rm ETC})$. 
The scale symmetry breaking in the ultraviolet region $\mu>\Lambda_{\rm TC}$ has no problem thanks to 
the naturalness as usual like in the QCD and the QCD-scale-up TC where the  theory has  only logarithmic divergences. 
Only possible source of the scale symmetry violation is from an effective theory for $\mu < m_F$.

Note first that since the effective Lagrangian ${\cal L}_{\rm inv}$ in Eq.(\ref{L:inv}) is scale-invariant, 
no mass corrections to $M_\phi$ arise from there. 
 The possible corrections thus come from the explicit breaking sector described 
by ${\cal L}_{S}$ and $V_\chi$ in Eqs.(\ref{Lag:S}) and (\ref{V:chi}).  
The TD potential $V_\chi$ includes terms up to quartic order of $\phi$, 
\begin{equation} 
  V_\chi = - \frac{1}{16}F_\phi^2 M_\phi^2 + \frac{1}{2} M_\phi^2 \phi^2 + \frac{4}{3} \frac{M_\phi^2}{F_\phi} \phi^3 
+ 2\frac{M_\phi^2}{F_\phi^2} \phi^4 + \cdots 
  \,, 
\end{equation} 
 from which we may evaluate the quadratically divergent correction to the TD mass at the one-loop level, 
arising from the quartic interaction of $\phi$: 
\begin{equation} 
  \delta M_\phi^2|_{\phi^4} \simeq \frac{m_F^2}{(4\pi)^2} \frac{24 M_\phi^2}{ F_\phi^2} 
  \,, \label{phi4:correction}
\end{equation} 
where we have regularized the quadratic divergence by the cutoff scale $m_F$. 
On the other hand, the Yukawa coupling terms in Eq.(\ref{Lag:S}) give similar corrections 
which is dominated by top-loop~\footnote{ 
Another source for the radiative breaking of scale symmetry might come from 
the techni-pion mass terms which would be included in 
the Lagrangian ${\cal L}_S$. 
 As was mentioned below Eq.(\ref{Lag:S}), however, 
 the techni-pion masses turn out to actually be higher than the cutoff $m_F$~\cite{Jia:2012kd}, 
 so that we can safely ignore them in evaluating radiative corrections based on 
 our effective Lagrangian. 
}:  
\begin{equation} 
  \delta M_\phi^2|_{\rm Yukawa} 
  \simeq 
-  \frac{m_F^2}{(4\pi)^2}  \frac{12 (3-\gamma_m)^2 m_t^2}{F_\phi^2} 
\,, \qquad 
\gamma_m  \simeq 1 
\,. \label{yukawa:correction} 
\end{equation}
  For a light TD with $M_\phi \simeq 125$ GeV,  
it turns out that   
the $\phi^4$ correction in Eq.(\ref{phi4:correction}) is 
suppressed by a factor of $(M_\phi/F_\phi)^2$ with the large TD decay constant $F_\phi$ (See Eq.(\ref{vals})), 
compared with the Yukawa correction in Eq.(\ref{yukawa:correction}).  
We may therefore evaluate the total $\delta M_\phi^2$ neglecting the $\phi^4$ correction.  
The quadratically divergent correction to the TD mass thus contribute to the bare mass $M_\phi^{(0)}\simeq  125$ GeV 
as follows:    
 \begin{equation} 
    M_\phi \simeq M_\phi^{(0)} \left[
  1 - \frac{3}{2 \pi^2} \frac{m_t^2 m_F^2}{M_\phi^2 F_\phi^2}
\right] 
\,. \label{Mphi:correction}
 \end{equation}

As it will turn out later, $(F_\phi M_\phi)$ is    
are related to $m_F$ involving $N_{\rm TC}$ and $N_{\rm TF}$ (See Eq.(\ref{Vac})). 
With the criticality condition (Eq.(\ref{criticality})), furthermore, 
we may write $N_{\rm TF} \simeq 4 N_{\rm TC}$~\cite{Appelquist:1996dq} and   
hence rewrite the correction term in Eq.(\ref{Mphi:correction}) to find  
\begin{eqnarray} 
 M_\phi 
 &\simeq& 
M_\phi^{(0)} \left[
  1 - \frac{1}{48 \kappa_V} \left( \frac{m_t}{m_F} \right)^2
\right] 
\nonumber \\ 
&\simeq &
 M_\phi^{(0)} \left[
 1 - \frac{0.025}{N_{\rm TC}}  
\right]
\,,
\end{eqnarray}
 for the one-family TC model with $N_D=4$, 
where in the second line we have used $m_F \simeq 319 \,{\rm GeV} \sqrt{\frac{4}{N_D}\frac{3}{N_{\rm TC}}}$ in 
Eq.(\ref{vals}) and 
$\kappa_V \simeq 0.7$ in Eq.(\ref{kappas}).  
Thus the one-loop radiative corrections give the shift by only about  
$(3/N_{\rm TC})$\% for the light TD with mass $M_\phi \simeq  125$ GeV, 
which is tiny enough to be natural against the quadratic divergence maximally breaking the scale symmetry. 
Higher loop corrections turn out to be even more dramatically suppressed by powers of  $(m_F/(4\pi F_\phi))^2$ 
due to the large TD decay constant $F_\phi$ (See Eq.(\ref{vals})).

\section{Techni-dilaton at LHC}  
 \label{TDLHC}

In this section we shall explore the TD discovery channels at LHC in comparison 
with the SM Higgs and the current ATLAS and CMS experimental data. 
We first estimate the size of the TD couplings  
by adopting results from a recent nonperturbative analysis of walking dynamics (Sec.~\ref{ETDC}). 
It then turns out that the TD in 1DM is invisible   
due to the highly suppressed couplings to the SM particles. 
The TD total width 
is evaluated for the 1FM models (Sec.~\ref{TW}). 
Taking the TD mass to be $125$ GeV and 600 GeV as the reference values, 
we then compute the LHC production cross sections times the branching ratios 
for the 1FMs with $N_{\rm TC}=3,5,7,9$,  
normalized to the corresponding quantities for the SM Higgs (Sec.~\ref{TDLHCsignals}). 
The TD signatures for the mass range $110 \, {\rm GeV} \le M_\phi \le 600$ GeV  
are compared with the presently reported experimental data (Sec.~\ref{comp-with-data}). 
Finally, the TD discovery signatures 
are discussed in detail (Secs.~\ref{125}).

\subsection{Estimate of the TD couplings} 
\label{ETDC}

The derived formulas for the TD couplings to the SM fermions and weak bosons,   
Eqs.(\ref{coupling:WW}), (\ref{coupling:ZZ}) and (\ref{yukawa:f:1}),    
imply a simple scaling between the TD couplings and the SM Higgs ones:  
\begin{eqnarray} 
  \frac{g_{\phi ff}}{g_{h_{\rm SM} ff}} 
&=& \frac{(3-\gamma_m) v_{\rm EW}}{F_\phi} 
\,, \qquad 
{\rm with} 
\qquad  
\gamma_m \simeq 1 
\,, \nonumber \\ 
\frac{g_{\phi WW/ZZ}}{g_{h_{\rm SM} WW/ZZ}} 
&=& \frac{ v_{\rm EW}}{F_\phi}
  \,. \label{g}
\end{eqnarray} 
On the other hand, the TD couplings to $gg$ and $\gamma\gamma$  
are given in Eqs.(\ref{TD:2gamma}) and (\ref{TDgg}) based on Eq.(\ref{TD:dip-dig}). 
In the case of a light TD such as the mass $M_\phi \sim $ 125 GeV,  
these couplings normalized to the corresponding quantities for the SM Higgs 
are approximately evaluated as 
\begin{eqnarray} 
\frac{g_{\phi gg}}{g_{h_{\rm SM} gg}} 
&\simeq & 
\frac{v_{\rm EW}}{F_\phi} 
 \Bigg| 
\frac{(3-\gamma_m) \beta^t_{\rm SM}(g_s) + \beta_F(g_s)}{ \beta_{\rm SM}^t(g_s)}
\Bigg|  
  \,,  \nonumber \\ 
\frac{g_{\phi \gamma\gamma}}{g_{h_{\rm SM} \gamma\gamma}} 
&\simeq & 
\frac{v_{\rm EW}}{F_\phi} 
 \Bigg| 
\frac{\beta_{\rm SM}^W(e) + (3-\gamma_m) \beta^t_{\rm SM}(e) + \beta_F(e)}{\beta_{\rm SM}^W(e) + \beta_{\rm SM}^t(e)} 
\Bigg|  
  \,, 
  \qquad  {\rm for} \quad \gamma_m \simeq 1
\,,  
\end{eqnarray} 
 where 
$\beta_{\rm SM}^t(g_s)=(2/3) \cdot g_s^3/(4\pi)^2$, $\beta_{\rm SM}^t(e) 
 = 3 \cdot (2/3)^2 \cdot 2/3 $ and  $\beta^W_{\rm SM}(e)= - 7/2 \cdot  e^3/(4\pi)^2$ 
including only the top quark and $W$ loop contributions at one-loop level. 
The above ratios are thus estimated by evaluating the techni-fermion contributions in $\beta_F$ 
once the model of WTC is fixed. 
For the 1DM and 1FM, we have 
\begin{eqnarray} 
  \beta_F(g_s)/(g_s^3/(4\pi)^2) &=& 
  \frac{2}{3} N_{\rm TC} \sum_Q N_Q 
  = 
  \Bigg\{ 
  \begin{array}{cc} 
    0 & {\rm for} \qquad  {\rm 1DM} \\ 
    \frac{4}{3} N_{\rm TC} 
& {\rm for} \qquad  {\rm 1FM}    
  \end{array}
 \,, \nonumber \\ 
    \beta_F(e)/(e^3/(4\pi)^2) &=& 
    \frac{2}{3} N_{\rm TC} \sum_F N_c^{(F)} Q_F^2 
    = 
  \Bigg\{ 
  \begin{array}{cc} 
    \frac{1}{3} N_{\rm TC} & {\rm for} \qquad  {\rm 1DM} \\ 
    \frac{16}{9} N_{\rm TC} & 
 {\rm for} \qquad  {\rm 1FM}    
  \end{array}
 \,,
\end{eqnarray}
where $N_c^{(F)}=1(3)$ for leptons (quarks). 
 Hence we find 
\begin{eqnarray} 
\frac{g_{\phi gg}}{g_{h_{\rm SM} gg}} 
&\sim& 
\frac{v_{\rm EW}}{F_\phi} 
\cdot 
 \Bigg\{  
 \begin{array}{cc} 
 4  & \qquad \textrm{for 1DM} \\
|2  +2 N_{\rm TC} | 
& \qquad \textrm{for 1FM} 
 \end{array} 
\,,  
\nonumber \\ 
\frac{g_{\phi \gamma\gamma}}{g_{h_{\rm SM} \gamma\gamma}} 
&\sim& 
\frac{v_{\rm EW}}{F_\phi} 
\cdot 
 \Bigg\{  
 \begin{array}{cc} 
|\frac{31}{47}  - \frac{9}{47} N_{\rm TC} | 
& \qquad \textrm{for 1DM} \\ 
| \frac{31}{47} - \frac{32}{47} N_{\rm TC} | 
& \qquad \textrm{for 1FM} 
 \end{array} 
\,.  \label{g-dip-dig}
\end{eqnarray}

The overall factor $(v_{\rm EW}/F_\phi)$ may be estimated through the PCDC relation Eq.(\ref{PCDC}): 
The TD decay constant $F_\phi$ and TD mass $M_\phi$ are related to the vacuum energy 
density ${\cal E}_{\rm vac}=\langle \theta_\mu^\mu\rangle/4$ through the PCDC relation 
as in Eq.(\ref{PCDC}): 
\begin{equation} 
 F_\phi^2 M_\phi^2 = - 4 \langle \theta_\mu^\mu \rangle 
\,.  \label{PCDC:rev} 
\end{equation} 
We may then write the vacuum energy density ${\cal E}_{\rm vac}$ 
in a generic manner as in Eq.(\ref{PCDC:gene}): 
\begin{equation} 
 \langle  \theta_\mu^\mu \rangle = 4 {\cal E}_{\rm vac}  
  = - \kappa_V \left( \frac{ N_{\rm TC} N_{\rm TF} }{2 \pi^2}  \right) m_F^4 
  \,,  
  \label{Vac}
\end{equation} 
where we have assumed that the techni-fermions belong to fundamental representation  for $SU(N_{\rm TC})$ gauge group and   
$\kappa_V$ is the overall coefficient in principle calculable by the  nonperturbative analysis. 
The dynamical techni-fermion mass $m_F$ can, on the other hand, be related to 
the techni-pion decay constant $F_\pi$:  
\begin{equation} 
  F_\pi^2 = \kappa_F^2 \frac{N_{\rm TC}}{4 \pi^2} m_F^2 
\,,   \label{PS}
\end{equation}
with the overall coefficient $\kappa_F$ and the property of $N_{\rm TC}$ scaling taken into account. 
The scale of $F_\pi$ is set by the electroweak scale $v_{\rm EW}$ along with $N_D$ as 
$F_{\pi} = v_{\rm EW}/\sqrt{N_D}$. 
With these combined, one can express 
$F_\phi M_\phi$ in Eq.(\ref{PCDC}) 
in terms of $N_{\rm TC}, N_{\rm TF}$ 
and $\kappa_{V,F}$, once $F_{\pi} = v_{\rm EW}/\sqrt{N_D}$ is fixed.

\begin{figure}

\begin{center} 
\includegraphics[scale=0.6]{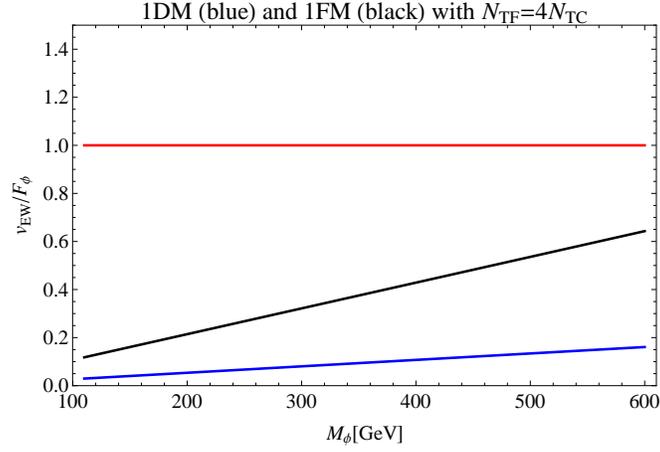}
\end{center}
\caption{ The plot of $(v_{\rm EW}/F_\phi)$   
as a function of the TD mass $M_\phi$ in a range from 110 to 600 GeV 
for the 1DM (blue line) and 1FM (black line) with $N_{\rm TF}=4 N_{\rm TC}$ fixed, 
in comparison with the SM Higgs case (red line).  
}  
\label{g-ratio-1FM}
\end{figure}

The values of $\kappa_V$ and $\kappa_F$ may be quoted from the latest result~\cite{Hashimoto:2010nw} on a 
ladder SD analysis for a modern version of 
WTC~\cite{Lane:1991qh,Appelquist:1996dq, Miransky:1996pd}~\footnote{
 In the previous work~\cite{Matsuzaki:2011ie} 
$\kappa_F$ and $\kappa_V$ were set to the values near the criticality, i.e., 
$(\kappa_F, \kappa_V) \simeq (1.5, 0.76)$, which is 
realized by taking the criticality limit $\alpha \to \alpha_c$ ($\Lambda_{\rm ETC}/m_F \to \infty$)~\cite{Hashimoto:2010nw}. 
The present paper has focused on an intermediate set of the values 
$(\kappa_F, \kappa_V) \simeq (1.4, 0.7)$ 
corresponding to a realistic situation $\Lambda_{\rm ETC}/m_F \simeq 10^3-10^4$ 
viable for the TC model-building. 
}: 
\begin{equation} 
 \kappa_V \simeq 0.7 \,, \qquad 
\kappa_F \simeq 1.4 
\,,  \label{kappas}
\end{equation} 
where $\kappa_F$ has been estimated based on the Pagels-Stokar formula~\cite{Pagels:1979hd}.  
In that case $N_{\rm TF}$ is fixed by the criticality condition~\footnote{
The estimated numbers based on the ladder approximation can have 
the uncertainties by about 30\%~\cite{Appelquist:1988yc}, which could result in 
uncertainty of 60\% for the diphoton event rate 
at $\simeq 125$ GeV, to be as large as about 2.7 times the SM Higgs case for $N_{\rm TC}=7$. 
} for the walking regime as~\cite{Appelquist:1996dq} 
\begin{equation} 
 N_{\rm TF} \simeq 4 N_{\rm TC}
 \,,
 \label{criticality}
\end{equation} 
where 
\begin{equation}
N_{\rm TF} = 2 N_D + N_{\rm EW-singlet} 
\,, 
\end{equation} 
with $N_{\rm EW-singlet}$ being the number of the  electroweak/color-singlet techni-fermions, ``dummy'' techni-fermions~\cite{Christensen:2005cb} 
introduced in order to fulfill the criticality condition, which serve to reduce the TD couplings  by enhancing  
$F_\phi$ through Eqs.(\ref{PCDC:rev}) and (\ref{Vac}). 
Note that $(v_{\rm EW}/F_\phi)$ is independent of $N_{\rm TC}$ when $N_{\rm TF} \simeq 4 N_{\rm TC}$ is used: 
\begin{eqnarray} 
\frac{v_{\rm EW}}{F_\phi}
&\simeq& 
\frac{1}{8 \sqrt{2} \pi} \sqrt{\frac{\kappa_F^4}{\kappa_V}} N_D \frac{M_\phi}{v_{\rm EW}}
\,. 
\end{eqnarray}
 Taking the original 1FM~\cite{Farhi:1980xs} with $N_{D} = 4$~as a definite benchmark, 
we thus evaluate $m_F$, $F_\phi$ and $(v_{\rm EW}/F_\phi)$
in Eq.(\ref{g}) to get 
\begin{eqnarray} 
m_F &\simeq& 319 \, {\rm GeV} \sqrt{ \frac{4}{N_D} \frac{3}{N_{\rm TC}}} 
\,, \nonumber \\ 
 F_\phi &\simeq& 1836 \, {\rm GeV} \left( \frac{4}{N_D} \right) \left( \frac{125\,{\rm GeV}}{M_\phi}  \right)  
\,, \nonumber \\  
\frac{v_{\rm EW}}{F_\pi}
&\simeq& 
0.134  \left( \frac{N_D}{4} \right) \left( \frac{M_\phi}{125\,{\rm GeV}} \right) 
\,. \label{vals}
\end{eqnarray}
  The plot of $(v_{\rm EW}/F_\phi)$ as a function of $M_\phi$ is  
shown in Fig.~\ref{g-ratio-1FM} for the 1DM and 1FM, in comparison with 
the SM Higgs.  
 In the case of 1DM, all the couplings are very small compared with the SM Higgs 
since the overall factor $(v_{\rm EW}/F_\phi)$ of the couplings  
in Eqs.(\ref{g}) and (\ref{g-dip-dig}) is of order 
${\cal O}(10^{-2})$ and other factors are not so large as 
to compensate the smallness.  
Thus the TD in the 1DM is invisible   
in all region.

As to the 1FM, the overall factor $(v_{\rm EW}/F_\phi)$ is four times larger than 
that of the 1DM, but is still small 
compared with the SM Higgs and hence the TD couplings to $WW, ZZ$ and $f\bar{f}$ in Eq.(\ref{g})
are substantially smaller than those of the SM Higgs.  
 On the other hand, the TD couplings to $gg$ and $\gamma\gamma$ in Eq.(\ref{g-dip-dig}) have 
 extra factors $|2 + 2 N_{\rm TC}|$ and $|(31 - 32 N_{\rm TC})/47|$ 
coming from techni-fermions as well as the $W$ and top quarks 
carrying the QCD color and electromagnetic charges. 
  The gluon fusion production thus becomes larger than the SM Higgs case 
  due to this extra factor. 
 Even considering this factor, 
the signals for $WW$, $ZZ$ and $f\bar{f}$ 
are extremely small compared with the SM Higgs, 
 unless we assume a gigantic number of $N_{\rm TC}$, roughly, $N_{\rm TC} > 50$.

However, the $\gamma\gamma$ event rate 
can be enhanced by the factors both from the $gg$ and $\gamma\gamma$ couplings, 
which can compensate 
the smallness of $(v_{\rm EW}/F_\phi)$ for a moderately large $N_{\rm TC}$:  
The $\gamma\gamma$ event rate 
may roughly be estimated as 
\begin{eqnarray} 
 R_{\gamma\gamma}^{(0)} 
& \equiv& 
 \left( \frac{g_{\phi gg}}{g_{h_{\rm SM} gg}}\right)^2  
 \cdot 
 \left( \frac{g_{\phi \gamma\gamma}}{g_{h_{\rm SM} \gamma\gamma}}\right)^2  
 \sim 
 \left( 0.134 \right)^4 \cdot 
 \left( 2  +2 N_{\rm TC} \right)^2 
 \left(\frac{31}{47} - \frac{32}{47} N_{\rm TC} \right)^2  
\,, \label{rough:dip:rate}
\end{eqnarray}
which exceeds unity when $N_{\rm TC} \ge 7$. 
More detailed estimation will be done later (See Table~\ref{tab:signal:125}, 
Figs.~\ref{Rdiphoton} and \ref{signal-strength}.).

\subsection{Total width} 
\label{TW}

  \begin{figure}

\begin{center} 
\includegraphics[scale=0.5]{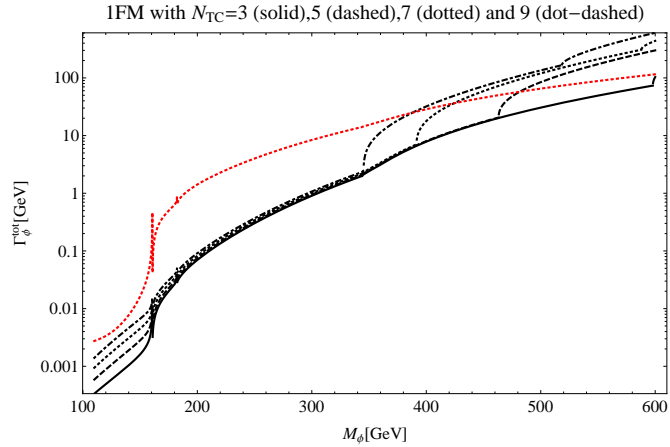}
\end{center}
\caption{ The TD total width as a function of the mass $M_\phi$ for the 1FMs with 
$N_{\rm TC}=3$ (solid), 5 (dashed), 7 (dotted) and 9 (dot-dashed), in comparison with the SM Higgs case (red-dashed curve). 
}  
\label{tot-width}
\end{figure}

 Using the values given in Eq.(\ref{vals}) 
we calculate the TD partial decay widths.  
In Appendix~\ref{widths} we  present those formulas relevant to the mass range 
$110 \, {\rm GeV} \le M_\phi \le 600$ GeV focusing on two-body decay modes.

Figure~\ref{tot-width} shows the TD total width as a function of the TD mass $M_\phi$ 
in the range $110\, {\rm GeV} \le M_\phi \le 600$ GeV 
for the 1FMs with $N_{\rm TC}=3,5,7,9$, 
in comparison with the SM Higgs case (red curve). 
The techni-fermion loops significantly contribute to  
the decays to $gg, \gamma\gamma, Z \gamma$, to make the total width larger.  
Such a high enhancement 
balances with the overall suppression of 
the TD couplings  as seen in Fig.~\ref{g-ratio-1FM} for the lower mass range 
up to around 500 GeV depending on $N_{\rm TC}$, to be comparable to  
the SM Higgs total width. 
Actually, above the mass around 500 GeV, 
a new two-body decay channel to color-triplet techni-pions $P_3^{\pm, 0}, P_3^{'}$ with the mass 
$m_{P_3} \simeq 299 \sqrt{3/N_{\rm TC}}$ GeV will be open 
to significantly enhance the total width. 
The total width will thus be much broader for the mass above 600 GeV,   
as seen in Fig.~\ref{tot-width}. 
The detailed analysis of techni-pions in the 1FM taking into account the walking features 
will be reported in another publication~\cite{Jia:2012kd}.

\subsection{TD LHC signatures} 
\label{TDLHCsignals}

As done in Refs.~\cite{Matsuzaki:2011ie,Matsuzaki:2012gd}, 
we shall define a ratio of the TD LHC production cross section times branching ratio 
normalized to the SM Higgs one: 
\begin{equation} 
    R_{X} 
    \equiv 
    \frac{ [\sigma_{\rm GF}(pp \to \phi) + \sigma_{\rm VBF}(pp \to \phi)]}{
[\sigma_{\rm GF}(pp \to h_{\rm SM}) + \sigma_{\rm VBF}(pp \to h_{\rm SM})]} 
\frac{BR(\phi \to X)}{BR(h_{\rm SM} \to X)}
\,,  \label{R:0}
\end{equation} 
where we have assumed that 
the dominant production cross section arises through gluon fusion (GF) and vector boson fusion (VBF) processes similarly to  
the SM Higgs case. 
Since the total width of TD is almost comparable with that of the SM Higgs up to the mass $\sim$ 600 GeV 
as seen from Fig.~\ref{tot-width},  
the narrow width approximation may be applicable.  
We may therefore 
rewrite the ratios of the production cross sections in terms of the ratios of the corresponding 
decay widths as~\cite{Georgi:1977gs}  
\begin{eqnarray} 
    \frac{\sigma_{\rm VBF}(pp \to \phi)}{\sigma_{\rm VBF}(pp \to h_{\rm SM})} 
  &=& 
\frac{\Gamma(\phi \to WW/ZZ)}{\Gamma(h_{\rm SM} \to WW/ZZ)} 
   \equiv r_{WW/ZZ} 
    \,, 
    \nonumber\\  
  \frac{\sigma_{\rm GF}(pp \to \phi)}{\sigma_{\rm GF}(pp \to h_{\rm SM})} 
  &=& \frac{\Gamma(\phi \to gg)}{\Gamma(h_{\rm SM} \to gg)} 
  \equiv r_{gg} 
  \,, 
  \label{r}
\end{eqnarray} 
which leads to 
\begin{equation}
R_{X} 
=
\left( 
\frac{\sigma_{\rm GF}(pp \to h_{\rm SM}) \cdot r_{gg}+\sigma_{\rm VBF}(pp \to h_{\rm SM})\cdot r_{WW/ZZ}}{\sigma_{\rm GF}(pp \to h_{\rm SM}) + \sigma_{\rm VBF}(pp \to h_{\rm SM})}
\right) 
r_{\rm BR}^X
\,, \label{R}
\end{equation}
where 
\begin{equation}
  r_{\rm BR}^X
  = 
  \frac{BR( \phi\to X)}{BR(h_{\rm SM} \to X)}
  \,. \label{rBR}
\end{equation} 
The SM Higgs branching ratios and LHC production cross sections at 7 TeV are read off 
from Ref.~\cite{Dittmaier:2011ti}.  
By using the formulas for the TD partial widths listed in Appendix~\ref{widths} together 
with the values of $m_F$, $F_\phi$ estimated from Eq.(\ref{vals}),  
the ratios $r_{WW/ZZ}, r_{gg}, r_{\rm BR}^X$ and $R_X$ in Eq.(\ref{R}) for the 1FMs 
are thus explicitly calculated as a function of $M_\phi$ only.

\begin{table}[t] 
\begin{tabular}{|c|c|c|} 
\hline 
\hspace{15pt} 1FM with $N_{\rm TC}$ \hspace{15pt} 
&\hspace{15pt} $r_{WW/ZZ}$ \hspace{15pt}
&\hspace{15pt} $r_{gg}$  \hspace{15pt}  \\ 
\hline \hline 
3   
& 0.018 & 1.2 \\ 
5  
& 0.018 & 2.7 \\ 
7   
& 0.018 & 4.8 \\ 
9 
& 0.018 & 7.6 \\ 
\hline 
\end{tabular}
\caption{
The estimated numbers at $M_\phi=125$ GeV relevant to the TD 7 TeV LHC production processes 
for the 1FMs, in comparison with the corresponding quantities for the SM Higgs. 
}\label{tab:production:125} 
\end{table} 
\begin{table}[t]  
\begin{tabular}{|c|c|c|} 
\hline 
\hspace{15pt} 1FM with $N_{\rm TC}$ \hspace{15pt} 
&\hspace{15pt} $r_{WW/ZZ}$ \hspace{15pt}
&\hspace{15pt} $r_{gg}$  \hspace{15pt}  \\ 
\hline \hline 
3 
& 0.41 & 12 \\ 
5 
& 0.41 & 27  \\ 
7  
& 0.41 & 48  \\ 
9  
& 0.41 & 74 \\ 
\hline 
\end{tabular}
\caption{
The same as in Table~\ref{tab:production:125} for $M_\phi=600$ GeV. 
}\label{tab:production:600} 
\end{table}

\subsubsection{Rate of production cross sections: $r_{gg}$ and $r_{WW/ZZ}$}

In Tables~\ref{tab:production:125} and \ref{tab:production:600}, 
 taking $M_\phi=125$ GeV and 600 GeV as the reference points, 
we make the lists for the estimated values of $r_{WW/ZZ}$ and $r_{gg}$  
for the 1FMs with $N_{\rm TC}=3,5,7,9$. 
   From these tables we see that 
the GF production cross sections 
 get enhanced because of the extra techni-quark loop contributions. 
 This becomes more operative when the TD mass gets larger, since 
the overall suppression of the TD coupling gets milder so that the coupling 
strength will be as much as the SM Higgs one, as seen from 
Fig.~\ref{g-ratio-1FM}.    
At the TD mass $M_\phi=600$ GeV, indeed, 
the GF productions are gigantically enhanced sensitively depending on $N_{\rm TC}$, 
while the VBF productions are suppressed simply 
due to the small TD couplings to the weak gauge bosons  
(See Table~\ref{tab:production:600}).

\subsubsection{Rate of branching fractions: $r_{\rm BR}^X$}

\begin{table}[t] 
\begin{tabular}{|c|c|c|c|c|} 
\hline 
\hspace{5pt} 
1FM with $N_{\rm TC}$ 
\hspace{5pt} 
&
\hspace{5pt}  
$r_{\rm BR}^{\gamma\gamma}$ 
\hspace{5pt} 
&
\hspace{5pt}  
$r_{\rm BR}^{gg}$ 
\hspace{5pt} 
&
\hspace{5pt} 
$r_{\rm BR}^{WW/ZZ}$  
\hspace{5pt} 
&
\hspace{5pt} 
$r_{\rm BR}^{b\bar{b}, c\bar{c}, \tau^+\tau^-}$  
\hspace{5pt} 
\\ 
\hline \hline 
3 & 0.095 & 7.9  
&0.12 & 0.47  \\ 
5 & 0.26 & 9.5  
& 0.065 & 0.26 \\ 
7 & 0.38 & 11 
&0.040 & 0.16  \\ 
9 & 0.46 & 11 
& 0.027 & 0.11 \\ 
\hline 
\end{tabular}
\caption{
The TD branching fraction at $M_\phi=125$ GeV for the 1FMs,  
normalized to the corresponding quantities for the SM Higgs.  
}\label{tab:BR:125} 
\end{table} 
\begin{table}[t] 
\begin{tabular}{|c|c|c|c|c|} 
\hline 
\hspace{5pt} 1FM with $N_{\rm TC}$ \hspace{5pt} 
&\hspace{5pt}  $r_{\rm BR}^{\gamma\gamma}$ \hspace{5pt} 
&\hspace{5pt}  $r_{\rm BR}^{gg}$ \hspace{5pt} 
&\hspace{5pt}  $r_{\rm BR}^{t\bar{t}}$ \hspace{5pt} 
&\hspace{5pt} $r_{\rm BR}^{WW/ZZ}$  \hspace{5pt} \\ 
\hline \hline 
3 & 323 & 14  
& 1.8 & 0.46 \\ 
5 & 277 & 10  
& 0.64 & 0.16  \\ 
7 & 354 & 13  
& 0.44 & 0.11  \\ 
9 & 414 & 14 
& 0.32 &0.079 \\ 
\hline 
\end{tabular}
\caption{
The same as in Table~\ref{tab:BR:125} for $M_\phi = 600$ GeV. 
}\label{tab:BR:600} 
\end{table}

The TD branching fractions for $M_\phi = 125$ GeV and 600 GeV  
normalized to the corresponding quantities for the SM Higgs (denoted as $r_{\rm BR}^X$) are 
shown in Tables~\ref{tab:BR:125} and \ref{tab:BR:600} 
for the 1FMs. 
 Note first that the branching fraction for decays to $WW^{(*)}$ and $ZZ^{(*)}$ are generically 
suppressed compared to the other channels. 
This is mainly because of their couplings, which are by about factor 2 
smaller than the couplings to fermions (See Eq.(\ref{g})) and are lack of extra 
factors developing with $N_{\rm TC}$ as in the couplings to $gg$ and $\gamma\gamma$ in Eq.(\ref{g-dip-dig}).

At the low mass $M_\phi=125$ GeV, 
the branching fractions to $WW^*, ZZ^*$, $b\bar{b}$ and $\gamma\gamma$ get suppressed 
compared to the SM Higgs case. 
This is  mainly due to the highly enhanced $gg$ decay rate by 
the extra factor $|2 + 2 N_{\rm TC}|^2$ coming from colored-fermion loop contributions as in Eq.(\ref{g-dip-dig}).

At the high mass $M_\phi=600$ GeV (Table~\ref{tab:BR:600}),  
 similarly to the low mass case, the branching fraction to $gg$ is enhanced 
 compared to the SM Higgs case, due to the extra colored-fermion loop contributions. 
 In contrast to the low mass case, the branching fraction to $\gamma\gamma$ is also enhanced at this high mass. 
This is because the contributions from techni-fermion loop overwhelm those from the $W$ loop at this high mass.     
 The branching fractions for $WW$, $ZZ$ and $t\bar{t}$ are suppressed 
 since the decays to $gg$ and $\gamma\gamma$ are enhanced. 
 The suppression of these branching fractions also come from another source: 
Actually, some decay channels to color-triplet techni-pions $P_3^{\pm, 0}$ and $P_3'$ become dominant 
above the threshold $2 m_{P_3} \simeq 600 \sqrt{3/N_{\rm TC}}$ GeV~\cite{Jia:2012kd}. 
(The rate of the branching fraction to $P_3$ pair is about 80\% at around 600 GeV, 
such that the decays to $WW$ and $ZZ$ as well as to $t\bar{t}$ 
become suppressed dramatically.) 
The threshold effect of decays to the $P_3$ pair  
 becomes eminent when $N_{\rm TC}$ is changed from 3 to 5 for $M_\phi =600$ GeV, 
so that the other decay amplitudes drop to be slightly suppressed. 
Such a threshold effect is milder (drop by about 10\%) for 
the $\gamma\gamma$ and $gg$ modes because of enhancement by techni-fermion loop contributions 
along with the number of $N_{\rm TC}$, while it is effective (drop by about 60\%) 
in the other decay modes fairly insensitive to the $N_{\rm TC}$ (See Table~\ref{tab:BR:600}).

 \subsubsection{Rate of event rates: $R_X$}

\begin{table}[t]  
\begin{tabular}{|c|c|c|c|c|c|} 
\hline 
\hspace{5pt} 
1FM with $N_{\rm TC}$ 
\hspace{5pt} 
&
\hspace{5pt}  
$R_{\gamma\gamma}$ 
\hspace{5pt} 
&
\hspace{5pt} 
$R_{gg}$ 
\hspace{5pt} 
&
\hspace{5pt} 
$R_{WW/ZZ}$  
\hspace{5pt} 
& 
$R_{b\bar{b}, c\bar{c}, \tau^+\tau^-}$ 
\\ 
\hline \hline 
3 & 0.11  
& 8.8 & 0.13 & 0.53 \\ 
5 & 0.64   
&25 & 0.16 & 0.66   \\ 
7 & 1.7  
& 48 & 0.18 & 0.72 \\ 
9 & 3.2  
& 79 & 0.19 & 0.75 \\ 
\hline 
\end{tabular}
\caption{
The TD signatures at $M_\phi=125$ GeV for the 1FMs, 
normalized to those of the SM Higgs. 
}\label{tab:signal:125} 
\end{table}  
\begin{table}[t] 
\begin{tabular}{|c|c|c|c|c|} 
\hline 
\hspace{5pt} 1FM with $N_{\rm TC}$ \hspace{5pt} 
&\hspace{5pt}  $R_{\gamma\gamma}$ \hspace{5pt} 
&\hspace{5pt}  $R_{gg}$ \hspace{5pt} 
&\hspace{5pt}  $R_{t\bar{t}}$ \hspace{5pt} 
&\hspace{5pt} $R_{WW/ZZ}$  \hspace{5pt} \\ 
\hline \hline 
3 & $3.4 \times 10^3$  
& $3.4 \times 10^3$ &  20  & 4.9 \\ 
5 & $6.4 \times 10^3$   
& $1.1 \times 10^4$ & 15  & 3.7 \\ 
7 & $1.4 \times 10^4$    
& $2.5 \times 10^4$ & 18  & 4.4\\ 
9 & $2.6 \times 10^4$  
& $4.3 \times 10^4$ & 20 & 5.0 \\ 
\hline 
\end{tabular}
\caption{
The same as in Table~\ref{tab:signal:125} for $M_\phi=600$ GeV. 
}\label{tab:signal:600} 
\end{table}

We now discuss the TD event rates normalized to the SM Higgs case, $R_X$ in Eq.(\ref{R}). 
In Tables~\ref{tab:signal:125} and \ref{tab:signal:600} we list the $R_X$ at 7 TeV LHC     
for $M_\phi=125$ GeV and 600 GeV in the case of the 1FMs. 
 At the high mass $M_\phi=600$ GeV, all the signatures 
are highly enhanced by the large GF production cross sections. 
 Though the decay rates to $WW$ and $ZZ$ are suppressed, 
these event rates $R_{WW}$ and $R_{ZZ}$ actually become larger than those of the SM Higgs 
due to the large enhancement of the GF production. 
Such enhanced $WW$ and $ZZ$ channels will thus be characteristic signatures of TD with the mass $\sim$ 600 GeV, 
to be tested by the upcoming 2012 data.  
Besides the enhanced $WW$ and $ZZ$ modes, 
the $\gamma\gamma$ modes at $M_\phi=$ 600 GeV become gigantically large 
as the number of $N_{\rm TC}$ gets 
increased (Table~\ref{tab:signal:600}), which yields a large cross section $\sim 1$ fb,  
to be testable at the LHC experiments~\cite{Matsuzaki:2011ie}~\footnote{
 In the previous analysis we adopted 
the universal scaling factor of the TD coupling from the SM Higgs 
$(3-\gamma_m) \simeq 2$, which turns out to be  
$(3-\gamma_m) \simeq 1$ for the TD couplings 
to the SM gauge bosons, since they are infrared-dominant quantities, 
as was mentioned in footnote~\ref{infra}. 
}.

For $M_\phi=$ 125 GeV,   
the diphoton channel gets enhanced as $N_{\rm TC}$ increases 
according to a simple scaling $R_{\gamma \gamma} \simeq R_{\gamma\gamma}^{(0)}$ in Eq.(\ref{rough:dip:rate}).  
On the other hand, other signatures such as $WW$, $ZZ$ and $f \bar{f}$  
are substantially suppressed simply due to the smallness of the overall 
factor of the couplings.  
Thus the light TD can be seen only through the diphoton channel as a large excess. 
Especially, the number of $R_{\gamma\gamma}$ at 125 GeV for the 1FM with $N_{\rm TC}=7$ 
coincides with the presently observed signal strength in the diphoton channel at the ATLAS and CMS 
experiments~\cite{:2012sk,CMS-PAS-HIG-12-001}, which we will explore more closely later.


 \subsection{Limits from the current LHC data} 
 \label{comp-with-data}

\begin{figure}[h] 

\begin{center} 
\includegraphics[scale=0.5]{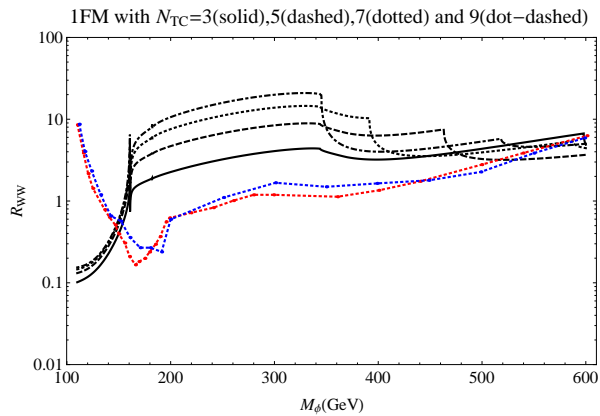}
\end{center}
\caption{ $R_{WW}$ as a function of the mass $M_\phi$ for the 1FMs with 
$N_{\rm TC}=3$ (solid), 5 (dashed), 7 (dotted) and 9 (dot-dashed), in comparison with the current ATLAS (red curve) and 
CMS (blue curve) 95\% CL upper limit~\cite{:2012sm,Chatrchyan:2012ty}. 
}  
\label{RWW}
\end{figure}

\begin{figure}[h] 

\begin{center} 
\includegraphics[scale=0.5]{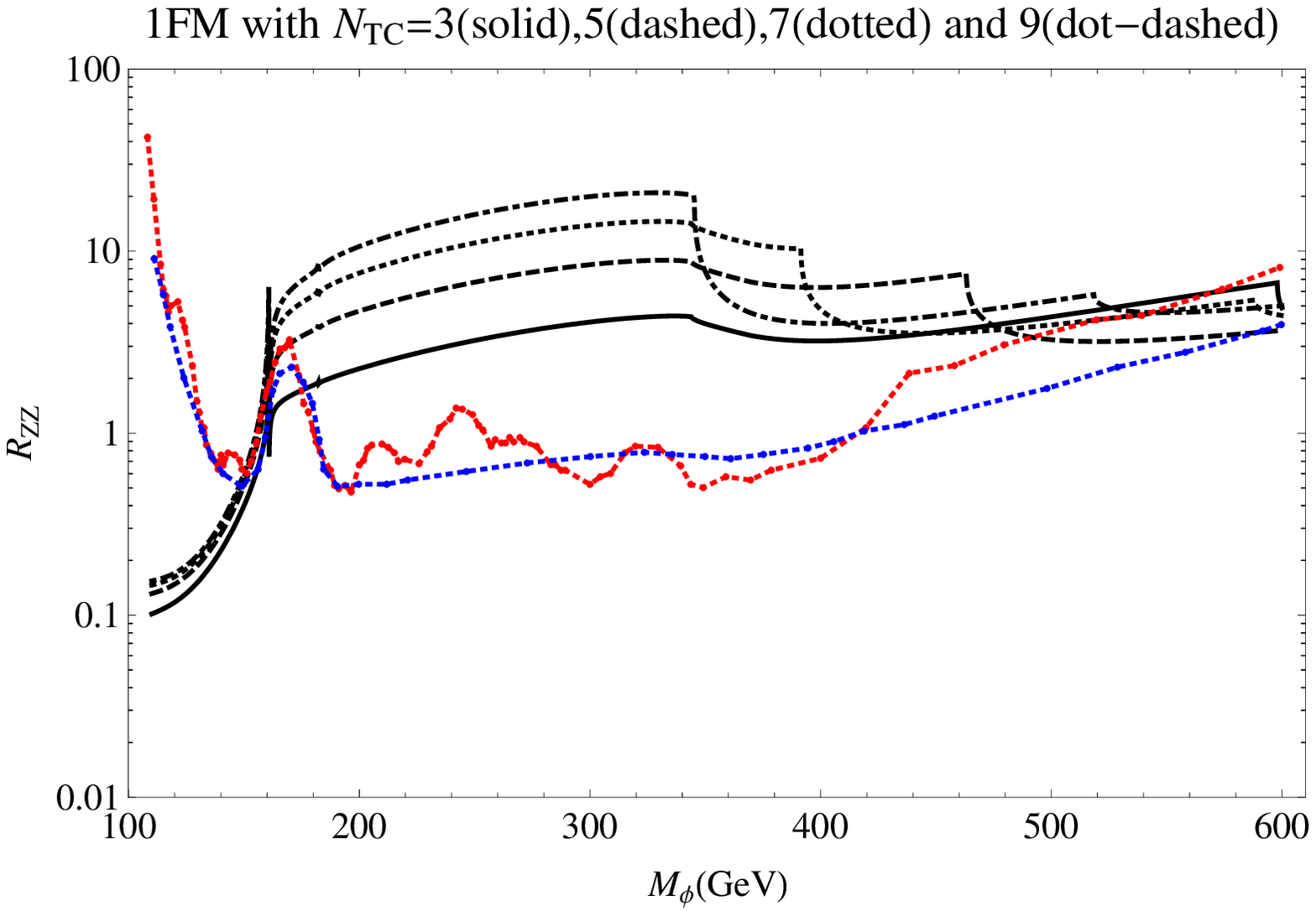}
\end{center}
\caption{ $R_{ZZ}$ as a function of the mass $M_\phi$ for the 1FMs with 
$N_{\rm TC}=3$ (solid), 5 (dashed), 7 (dotted) and 9 (dot-dashed), in comparison with the current ATLAS (red curve) and 
CMS (blue curve) 95\% CL upper limit~\cite{:2012sm,Chatrchyan:2012ty}. 
}  
\label{RZZ}
\end{figure}

In Figs.~\ref{RWW} and \ref{RZZ} 
we show the comparison with the current 95\% CL upper limits 
on the ratios $R_{WW}$ and $R_{ZZ}$ from the ATLAS and CMS experiments~\cite{:2012sm,Chatrchyan:2012ty}.  
We see that 
the current data on the $WW$ and $ZZ$ channels exclude the TD mass in the mass range 
$145 \, {\rm GeV} \lesssim  M_\phi \lesssim 600$ GeV. 
The bumps at around 500 GeV 
appear because 
the decay channels to color-triplet techni-pions start to open depending on $N_{\rm TC}$ 
to be dominant~\cite{Jia:2012kd}.

\begin{figure}[h] 

\begin{center} 
\includegraphics[scale=0.5]{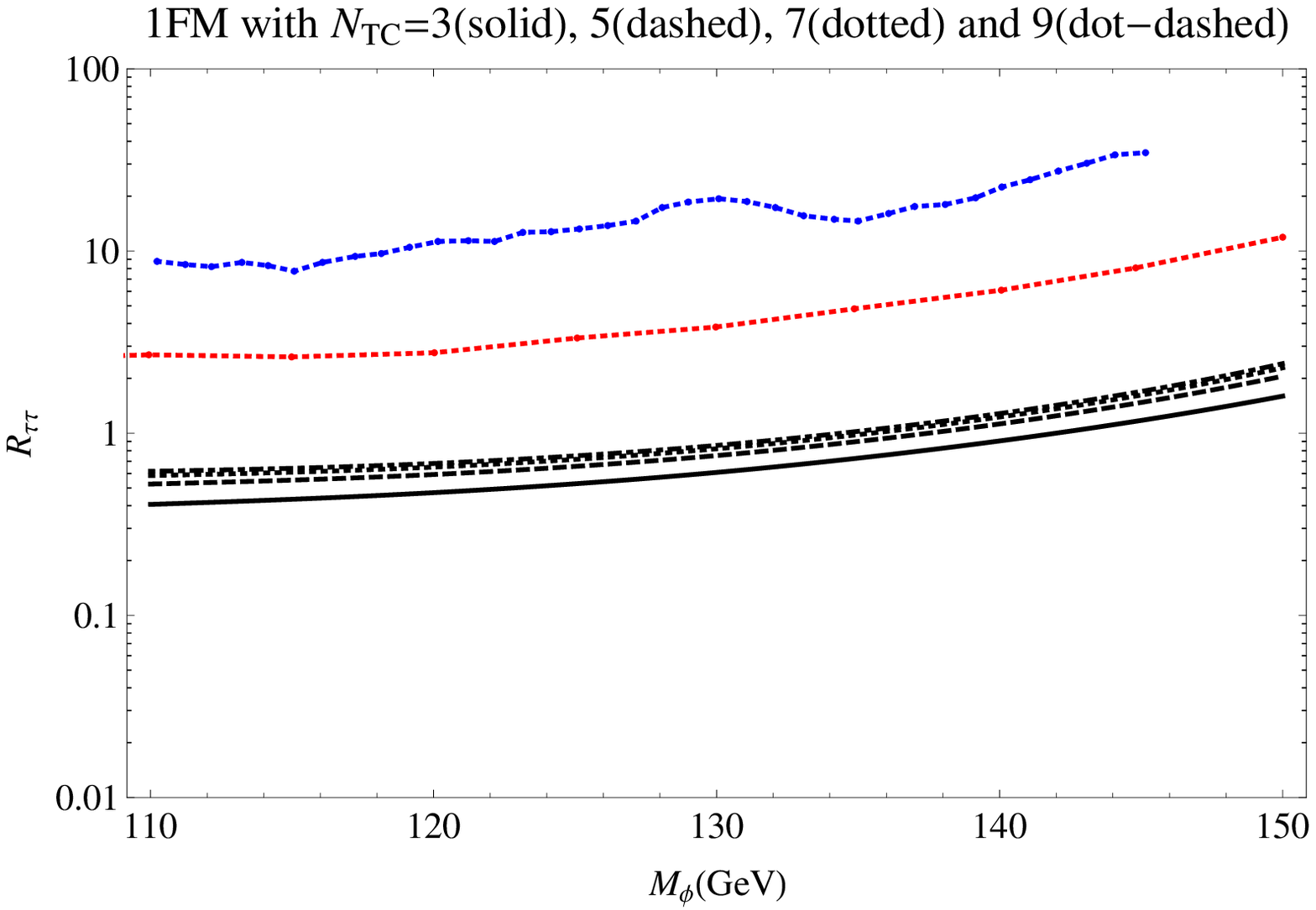}
\end{center}
\caption{ $R_{\tau^+\tau^-}$ as a function of the mass $M_\phi$ for the 1FMs with 
$N_{\rm TC}=3$ (solid), 5 (dashed), 7 (dotted) and 9 (dot-dashed), 
in comparison with the current ATLAS (red curve) and 
CMS (blue curve) 95\% CL upper limit~\cite{ATLAS-CONF-2012-014}. 
}  
\label{Rtautau}
\end{figure}

\begin{figure}[h] 

\begin{center} 
\includegraphics[scale=0.5]{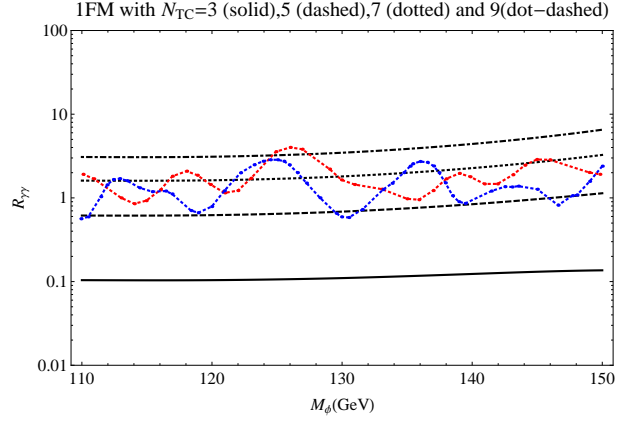}
\end{center}
\caption{ $R_{\gamma\gamma}$ as a function of the mass $M_\phi$ for the 1FMs with 
$N_{\rm TC}=3$ (solid), 5 (dashed), 7 (dotted) and 9 (dot-dashed)
in comparison with the current ATLAS (red curve) and 
CMS (blue curve) 95\% CL upper limit~\cite{:2012sk,CMS-PAS-HIG-12-001}. 
}  
\label{Rdiphoton}
\end{figure}

In Figs.~\ref{Rtautau} and \ref{Rdiphoton} the TD signatures in 
the $\tau^+ \tau^-$ and $\gamma\gamma$ channels are compared with 
the presently reported experimental data for the low mass region below 150 GeV~\cite{:2012sk,ATLAS-CONF-2012-014}.  
The $\tau^+\tau^-$ signatures are much below the upper limits,  
due to the large suppression of the TD coupling (See the reference point $M_\phi=125$ GeV in Fig.~\ref{g-ratio-1FM} or 
Table~\ref{tab:signal:125}.) 
Remarkably, the $\gamma\gamma$ signatures are close to the observed data 
as $N_{\rm TC}$ is increased, to coincide at around 125 GeV 
when $N_{\rm TC}=7$ as in Table~\ref{tab:signal:125}. 
We will address the 125 GeV TD in detail below.

\subsection{TD discovery signatures at 125 GeV} 
\label{125}

We shall look into the 125 GeV TD more closely through 
the predicted signals in the diphoton and weak boson channels. 
As seen in Table~\ref{tab:signal:125} and Fig.~\ref{Rdiphoton}, 
the TD diphoton signals are fairly sensitive to the number of $N_{\rm TC}$:  
When $N_{\rm TC}=7$ it is close to the amount of the presently observed signal strength $\sim 2 \times \sigma_{\rm h_{\rm SM}} 
\times {\rm BR}(h_{\rm SM} \to \gamma\gamma)$~\cite{:2012sk,CMS-PAS-HIG-12-001}, while 
it exceeds the present observation for $N_{\rm TC}\ge 8$. 
This feature can be understood by considering a ratio $R_{\gamma\gamma}/R_{WW/ZZ}$ whose 
$N_{\rm TC}$-dependence can be roughly described at $M_\phi=125$ GeV:  
\begin{equation} 
\frac{R_{ \gamma\gamma}}{R_{WW/ZZ}} \Bigg|_{N_{\rm TC}} 
\simeq 
\frac{r_{\rm BR}^{\gamma\gamma}}{r_{\rm BR}^{WW/ZZ}}
\simeq 
\Bigg| \frac{31}{47} - \frac{32}{47} N_{\rm TC} \Bigg|^2
\,, 
\end{equation}   
 which follows Eq.(\ref{rough:dip:rate}). 
The diphoton excess therefore grows even more as $N_{\rm TC}$ is increased.  
It is sharply contrasted to other channels 
including the $WW/ZZ$ and fermionic channels which are almost insensitive 
to $N_{\rm TC}$, staying in the range consistent with the present data~\cite{:2012sm,Chatrchyan:2012ty,ATLAS-CONF-2012-014} 
as seen from Figs.~\ref{RWW}, \ref{RZZ} and \ref{Rtautau}.

 \begin{figure}[h] 

\begin{center} 
\includegraphics[width=5.5cm]{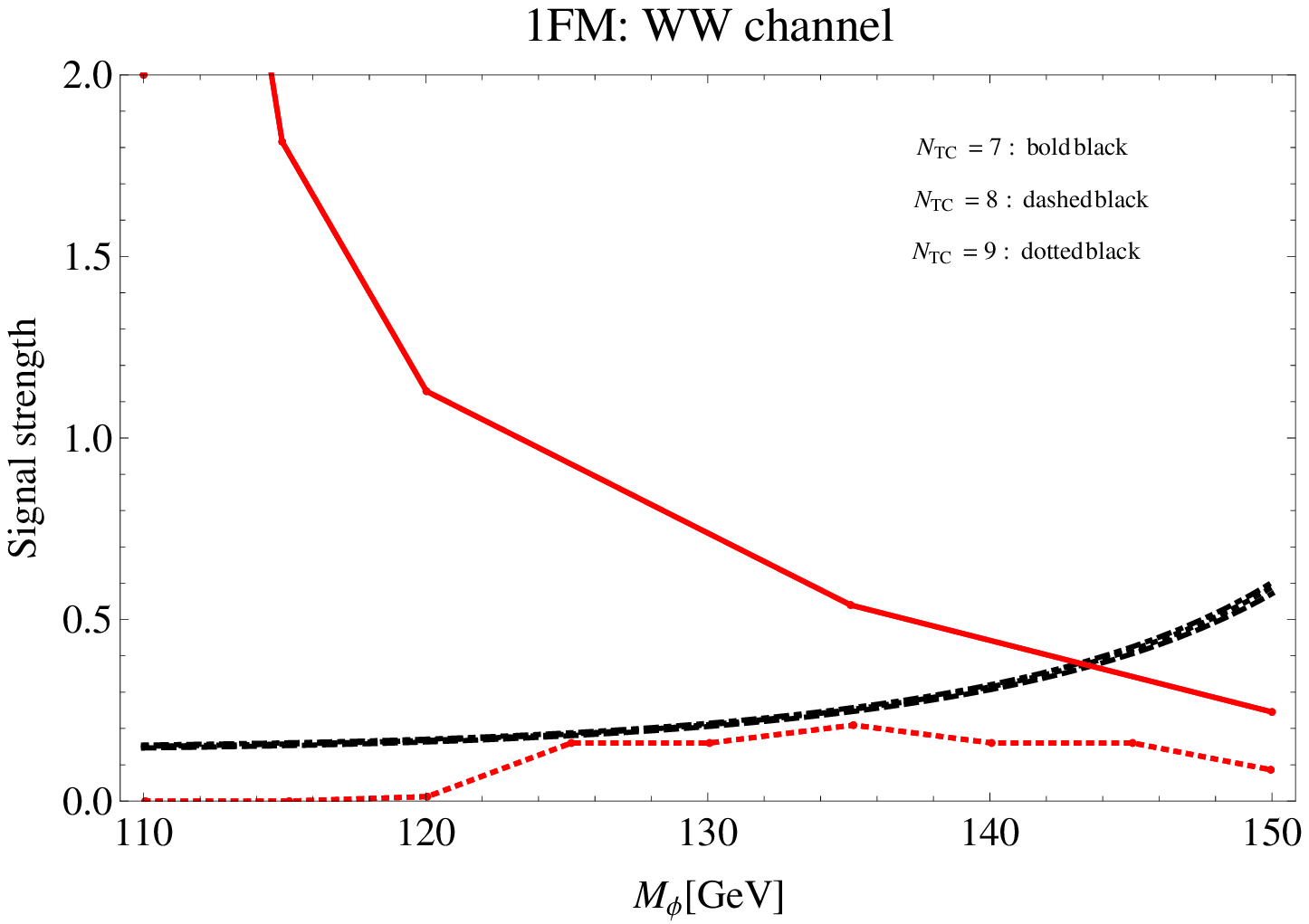}
\includegraphics[width=5.5cm]{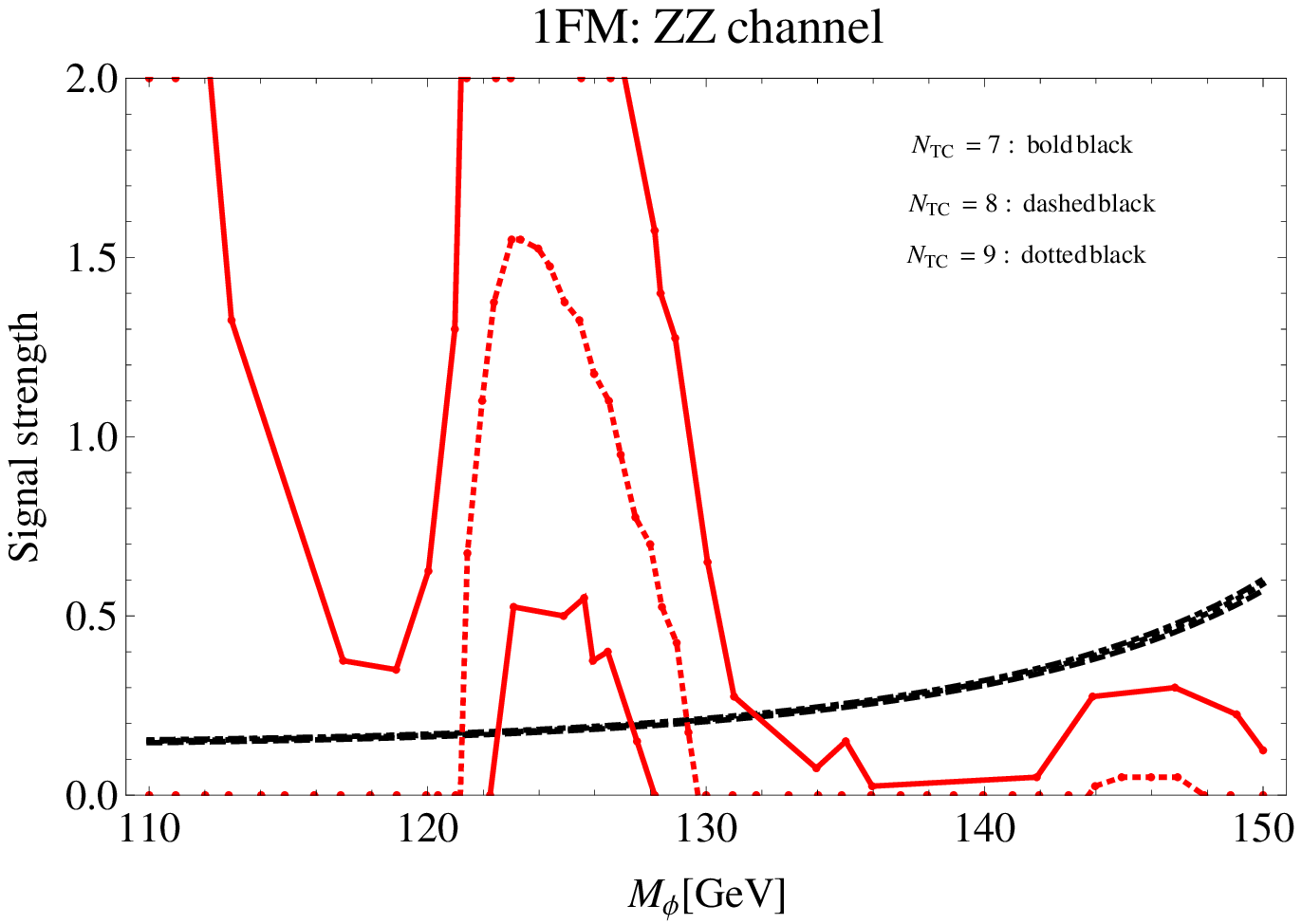}
\includegraphics[width=5.5cm]{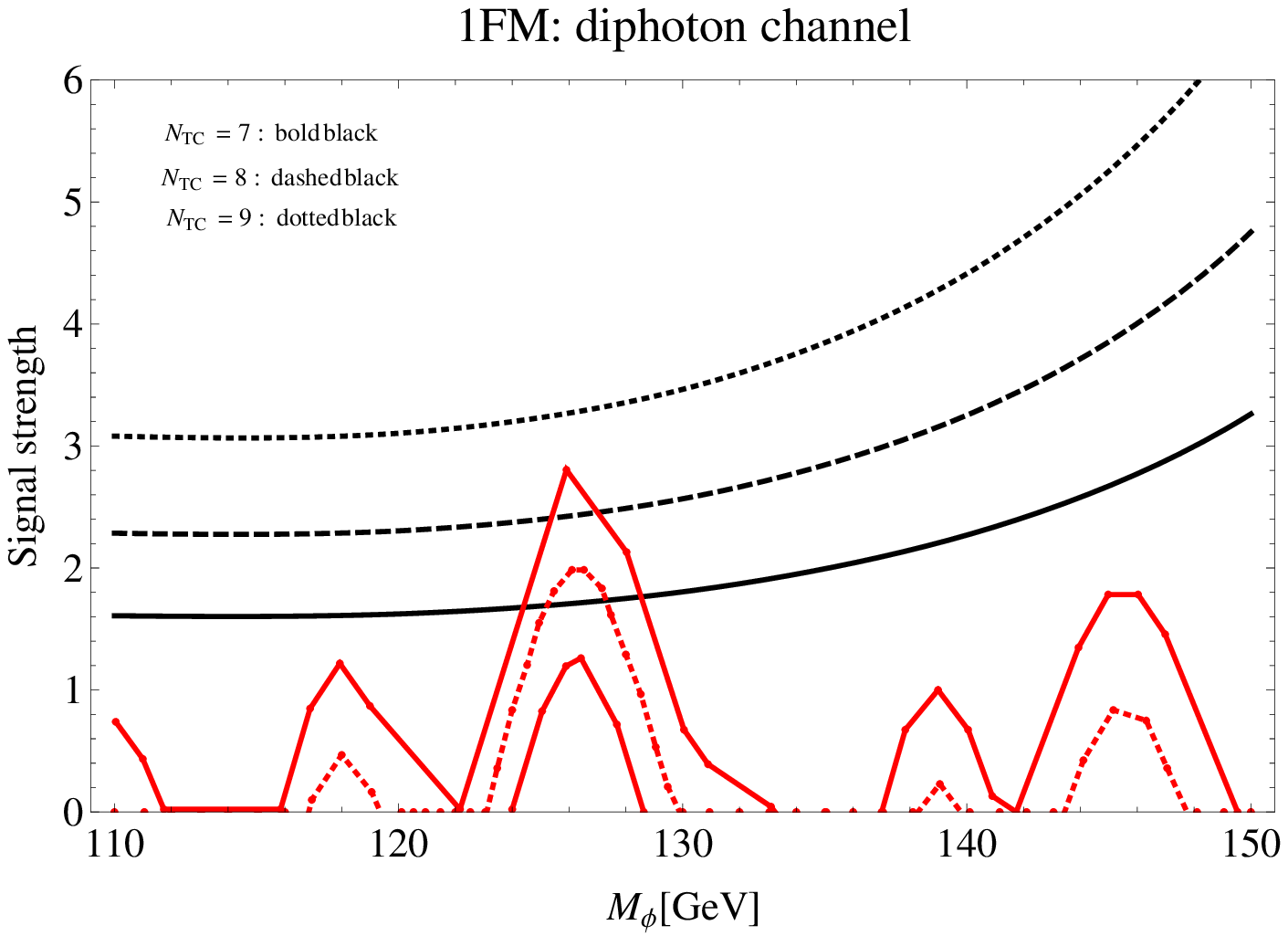}
\end{center}
\caption{ 
The plots of $R_{WW}, R_{ZZ}$ and $R_{\gamma\gamma}$ in the low mass range $110 \,{\rm GeV }\le M_\phi \le 150\, {\rm GeV}$
for the 1FMs with $N_{\rm TC}=7, 8, 9$ (black-solid, dashed and dotted curves), 
in comparison with the best-fit signal strengths estimated by the ATLAS experiment~\cite{ATLAS-CONF-2012-019} 
(red-dashed curves) including the 1 $\sigma$ uncertainty band (denoted by red-solid curves) read from the reference. 
The possible theoretical uncertainty about 30\% described in the text has been incorporated 
in the respective black thin curves. 
}  
\label{signal-strength}
\end{figure}

In Fig.~\ref{signal-strength} we also plot the TD signal strengths in the weak boson and diphoton channels 
in the case of the 1FMs with $N_{\rm TC}=7,8,9$ 
for the low mass range $110 \,{\rm GeV }\le M_\phi \le 150\, {\rm GeV}$, 
in comparison with the best-fit signal strengths estimated by the ATLAS experiment~\cite{ATLAS-CONF-2012-019} 
including the 1 $\sigma$ uncertainty band (denoted by red-solid curves) read from the reference. 
  Fig.~\ref{signal-strength} indeed tells us that 
when $N_{\rm TC}=7- 9$ the TD signals are consistent with 
the presently observed signal strengths in the weak boson and diphoton channels.

 Besides the boson channels, 
the predicted signals in the fermionic channels  
such as the decay channels to $\tau^+ \tau^-$~\cite{ATLAS-CONF-2012-014} 
as well as to $b \bar{b}$~\cite{ATLAS-CONF-2012-015} are consistent as well: 
As seen from Fig.~\ref{Rtautau} the signatures at around 125 GeV 
in the fermionic channels get suppressed compared to the SM Higgs case, 
mainly due to relative enhancement in the $gg$ decay mode (See Table~\ref{tab:BR:125})~\footnote{
Hence the TD may not account for another excess about 2 $\sigma$  
in the $b \bar{b}$ channel observed at Tevatron~\cite{TEVNPH:2012ab}. }. 
Such suppressed signals turn out to be much below the presently reported 
95\% CL upper limits~\cite{ATLAS-CONF-2012-014,ATLAS-CONF-2012-015}, to be 
consistent with the best-fit signal strengths for the $\tau^+ \tau^-$ and $b \bar{b}$ modes within 
the large systematic uncertainties at around 125 GeV~\cite{ATLAS-CONF-2012-019,CMS-PAS-HIG-12-008}.

Thus, if the excessive diphoton signals develop at the upcoming experiments to reach 
the desired significance level, while other channels essentially stay at  
the present significance, it  would imply  
the discovery of the 125 GeV TD. 
The excess at around 125 GeV only in the diphoton channel 
will be a salient feature of the TD discriminated from the SM Higgs~\cite{Matsuzaki:2012gd}.

A global analysis of experimental constraints on Higgs-like objects at around 125 GeV 
has recently been discussed by several authors~\cite{Azatov:2012bz}, 
where the size of deviation from the SM Higgs couplings as in Eq.(\ref{g}) 
is treated as a free parameter.    
Those analyses were, however, done by assuming that there is no contribution to couplings to $gg$ and $\gamma\gamma$ 
from the sector beyond the SM, or QED and QCD are fully embedded into a single scale-invariant/conformal field theory, 
which allows us to evaluate the couplings to $gg$ and $\gamma\gamma$ in terms of known contributions from the SM particles 
as done in the literatures regarding other dilaton scenario~\cite{Clark:1986gx,Goldberger:2007zk,Campbell:2011iw}. 
Note that both analyses cannot  be applied to the 125 GeV TD in the WTC scenario  
where the WTC contributions are incorporated in  
the TD couplings to $gg$ and $\gamma\gamma$, completely separated 
from the SM sector contributions (See Eqs.(\ref{TD:2gamma}) and (\ref{TDgg})).

\section{Summary}  
 \label{summary}

 In summary, we have explored in detail the TD signatures at LHC as an extension from 
the previously reported papers~\cite{Matsuzaki:2011ie,Matsuzaki:2012gd}. 
We first addressed that the TD couplings to techni-fermions are   
derived based on the Ward-Takahashi identity for the dilatation current coupled to TD.  
It was clarified that all the couplings to the SM particles are induced from techni-fermion loops: 
The Yukawa couplings to the SM fermions arise due to  
ETC-induced four-fermion interactions reflecting the ultraviolet feature of WTC 
characterized by the anomalous dimension $\gamma_m \simeq 1 $.  
The couplings to the SM gauge bosons, on the other hand, are determined 
 by the infrared features fixed solely by the low-energy theorem.

We also refined the low-energy effective Lagrangian for TD in a way consistent 
with the Ward-Takahashi identities mentioned above. 
The Lagrangian was 
based on the nonlinear realization of both the scale and chiral symmetries, where 
the scale invariance is ensured by including a spurion field which reflects 
the explicit breaking induced from the dynamical generation of techni-fermion mass itself. 
 We further showed that the light TD mass is stable to be natural against radiative corrections 
breaking the scale symmetry which would arise from outside of the walking regime.

The estimate of the TD couplings was done by using the recent result of 
the ladder SD analysis together with the PCDC relation. 
 For the 1DMs, the overall factor $(v_{\rm EW}/F_\phi)$ of the TD couplings 
 is so small that all the signatures are invisible at the LHC. 
As to the 1FMs, the event rates for $WW$, $ZZ$ and $f\bar{f}$ 
are small compared with the SM Higgs due to the smallness of 
the couplings. 
On the other hand, the $\gamma\gamma$ event rate becomes enhanced  
due to the two enhancement factors from both 
the $gg$ and $\gamma\gamma$ couplings (See Eq.(\ref{g-dip-dig})).

We then discussed the TD LHC signatures in the 1FMs with various $N_{\rm TC}$ 
 for the TD mass 110 GeV - 600 GeV,  
in comparison with the SM Higgs. 
It turned out that the light TD can be discovered as a large excess  
relative to the SM Higgs 
at around 125 GeV   
only in the diphoton channel: 
If the currently observed diphoton excess could come mainly from 
the VBF production,  
the 125 GeV TD would be excluded,    
which is to be soon tested in the upcoming 2012 LHC data. 
\\

{\bf Note added}: 
 Very recently, on July 4th, 2012, 
 the ATLAS and CMS have reported 5 sigma discovery of a Higgs-like particle 
at around 125 GeV, particularly in the diphoton and $ZZ^{*} \to 4l$ channels~\cite{070412}. 
 While the diphoton signal is consistent with the TD prediction, the $ZZ^*$ 
  appears to be somewhat larger than the TD prediction in the text (See Eq.(\ref{tab:signal:125})). 
However, the TD can still be consistent with the new data in the following way.

 The values presented in Table~\ref{tab:signal:125} are actually only typical ones based on the 
ladder approximation which are subject to certain uncertainties up to 30\%  
observed for the critical coupling and hadron spectrum in QCD~\cite{Appelquist:1988yc,Aoki:1990yp,Harada:2004qn}.   
 We may include this 30\% uncertainty in estimation of $F_\phi$ in Eq.(\ref{vals}) 
 for each independent factor $\kappa_V$ (Eq.(\ref{Vac})), $\kappa_F^2$ (Eq.(\ref{PS}))  
and the criticality condition $N_{\rm TF}/(4N_{\rm TC})$ (Eq.(\ref{criticality})). 
We then find the total size of uncertainties on $F_\phi$  
to be about 60\%. This implies a shift of $F_\phi$: 
$F_\phi \simeq 1836 \, {\rm GeV} \sqrt{4/N_D} \to$ as low as 
$\simeq 700 \, {\rm GeV}\sqrt{4/N_D}$. 
Thus the overall ratio of the TD coupling to the SM Higgs case could be  
$(v_{\rm EW}/F_\phi) \simeq 0.3$, compared with the typical value in the text, 
$(v_{\rm EW}/F_\phi) \simeq 0.134$ in Eq.(\ref{vals}).  
 This ratio $\simeq 0.3$ 
is large enough 
to be compensated by the enhanced GF production (See Eq.(\ref{g-dip-dig})) to yield 
the $ZZ^*$ signal of the 125 GeV TD comparable to that of the SM Higgs. 
(Such uncertainties of the ladder approximation will be settled by 
more reliable nonperturbative calculations such as lattice simulations. )

 One might wonder if this modification would enhance the TD modes to the 
 SM fermions roughly 5 times $(\simeq (0.3/0.134)^2)$ larger than the values in Table~\ref{tab:signal:125}, 
to be unacceptable experimental upper limits~\cite{070412}. 
Actually, this is not the case if we take into account the proper mass generation for 
the third-generation SM fermions:  
In Eq.(\ref{g}) we have adopted 
the simple-minded ETC scheme for all the SM fermion masses 
 through the techni-fermion condensate with the anomalous dimension $\gamma_m \simeq 1$ 
($3-\gamma_m \simeq 2$). 
In that case, however, 
the heavy third-generation quarks and leptons may not get a sufficient 
amount of contributions and still be lighter than the realistic ones.  
It was found that  if we include additional four-fermion interactions like strong ETC, 
the anomalous dimension becomes much larger $1<\gamma_m<2$,  
which can boost the ETC-origin mass to arbitrarily large up till  the techni-fermion mass scale 
(``strong ETC model'')~\cite{Miransky:1988gk,Matumoto:1989hf}.  
To accommodate the realistic fermion mass generation,  
it may thus be reasonable to put $\gamma_m \simeq 2$ in Eq.(\ref{g}) 
for the TD coupling to the third-generation fermions such as 
$\tau^+\tau^-$ and $b \bar{b}$. 
 Then the overall factor $(3-\gamma_m)$ of the TD Yukawa couplings 
to $b\bar{b}$ and $\tau^+\tau^-$ in Eq.(\ref{g}) becomes unity, 
 $(3-\gamma_m) \simeq 1$, which almost 
compensates the factor 5 shift of $(v_{\rm EW}/F_\phi)^2$
in the above modification.

With these prescriptions made, 
the TD signatures at 125 GeV now slightly become modified 
from those listed in Table~\ref{tab:signal:125}. 
We show the modified event rates $R_{X}$ of the 125 GeV TD 
in Table~\ref{tab:signal:125:modi}, taking $F_\phi = 700, 800, 900, 1000$ GeV 
as the reference points for $N_{\rm TC}=4,5$ (instead of $N_{\rm TC} \ge 7$ in the text). 
  The numbers shown in Table~\ref{tab:signal:125:modi} indeed imply that 
 the TD signatures at around 125 GeV are consistent with the 
best-fit signal strengths ($\mu = \sigma/\sigma_{\rm SM}$) reported by the ATLAS and CMS  
experiments~\cite{070412}. 
 In particular, the diphoton signal about twice larger than the expectation from the SM Higgs 
can be explained by the TD, which is due to the enhanced GF production cross section.  
This enhanced GF production is in contrast to the VBF production which is suppressed for the TD:  
The TD signal strength in the diphoton plus dijet channel tends to be 
smaller than the standard model Higgs prediction, 
simply because of the suppression of the overall TD coupling 
compared to the SM Higgs. 
Similar suppressions are also seen in other exclusive channels like 
($2l 2 \nu +  2j$) and ($\tau^+  \tau^- + 2j$) as well as $b \bar{b}$ 
originated from the vector boson fusion and vector boson associate productions. 
This salient feature will be tested to be confirmed or excluded by the upcoming 2012 data.

\begin{table} 
\begin{tabular}{|c|c||c|c|c|} 
\hline 
\hspace{10pt} $F_\phi$ [GeV] \hspace{10pt} & 
\hspace{10pt} $N_{\rm TC}$ \hspace{10pt} & 
\hspace{10pt} $R_{b\bar{b},\tau^+\tau^-}$ \hspace{10pt} & 
\hspace{10pt} $R_{WW/ZZ}$ \hspace{10pt} &
\hspace{10pt}  $R_{\gamma\gamma}$ \hspace{10pt}  \\ 
\hline \hline   
700 & 4 & 1.3 & 1.3 & 1.7 \\ 
    & 5 & 1.3 & 1.3 & 3.8 \\ 
\hline 
800 & 4 & 0.97 & 0.97 & 1.3 \\ 
    & 5 & 1.0 & 1.0 & 2.9 \\ 
\hline 
900 & 4 & 0.77 & 0.77 & 1.1 \\ 
    & 5 & 0.79 & 0.79 & 2.3 \\ 
\hline 
1000 & 4 & 0.62 & 0.62 & 0.85 \\ 
     & 5 & 0.64 & 0.64 & 1.9 \\    
\hline  
\end{tabular}
\caption{The modified TD signatures at 125 GeV taking into account the prescriptions described in the text. } 
\label{tab:signal:125:modi}
\end{table}

 After submitting the paper, we have posted on arXiv a paper (arXiv:1207.5911) 
performing a goodness-of-fit of the 125 GeV TD signal based on the latest data 
(as of July 25, 2012) of the LHC by extending the above reanalysis: The TD 
actually turns out to be favored by the current LHC data, slightly better than 
the SM Higgs. 
Also related papers appeared after the submission~\cite{Lawrance:2012cg}.

\section*{Acknowledgments}

We would like to thank M.~Harada for useful comments and discussions. 
This work was supported by 
the JSPS Grant-in-Aid for Scientific Research (S) \#22224003 and (C) \#23540300 (K.Y.).

\appendix 
\renewcommand\theequation{\Alph{section}.\arabic{equation}}

\section{The partial decay widths} 
\label{widths}

In this appendix we shall present the formulas for the TD partial decay widths 
focusing on the two-body decays to the SM particles.

\begin{itemize} 

\item $\phi \to f\bar{f}$:    

\begin{equation} 
\Gamma(\phi \to f \bar{f}) 
= 
\frac{ (3-\gamma_m)^2 N_c^{(f)} m_f^2 M_{\phi}}{8 \pi F_{\phi}^2} 
\left( 1 - \tau_f \right)^{3/2}
\,, 
\qquad 
\tau_f = \frac{4m_f^2}{M_\phi^2} 
\,, \label{TD:ff}
\end{equation}
 where $N_c^{(f)}=3(1)$ for quarks (leptons).

\item $\phi \to WW^*, ZZ^*$:

\begin{eqnarray} 
\Gamma(\phi \to WW^*) 
&=& 
\delta_{W^*} \frac{3  G_F m_W^4 M_\phi}{16 \sqrt{2} \pi^3 F_\phi^2} 
 R\left(\frac{m_W^2}{M_\phi^2}\right) 
\,, 
\nonumber \\ 
\Gamma(\phi \to ZZ^*) 
&=& 
\delta_{Z^*} \frac{3  G_F m_Z^4 M_\phi}{16 \sqrt{2} \pi^3 F_\phi^2} 
 R\left(\frac{m_Z^2}{M_\phi^2}\right) 
\,, 
\nonumber \\  
R(x) &=& 
\frac{3 (1- 8 x + 20 x^2)}{\sqrt{4 x-1}} \cos^{-1} \left( \frac{3 x-1}{2 x^{3/2}} \right) 
- \frac{(1-x) (2 - 13 x + 47 x^2)}{2 x} 
- \frac{3}{2} (1 - 6 x + 4 x^2) \log  x 
\,, 
\nonumber \\ 
\delta_{V^*} &=& 
\Bigg\{ 
\begin{array}{cc} 
1  & \qquad {\rm for} \qquad W  
\\ 
\frac{7}{12} - \frac{10}{9} s_W^2 + \frac{40}{27} s_W^4 & 
\qquad {\rm for} \qquad Z  
\end{array} 
\,,  
\end{eqnarray} 
 where $G_F$ is the Fermi coupling constant defined as $G_F/\sqrt{2} = g_W^2/(8 m_W^2)$, and 
$s_W(c_W)$ denotes the weak mixing angle defined as $s_W=e/g_W$ $(c_W=e/g_Y)$ with the electromagnetic (EM) coupling $e$ and 
$SU(2)_W$ ($U(1)_Y$) gauge coupling $g_W$ ($g_Y$).

\item $\phi \to \gamma\gamma$: 

\begin{eqnarray} 
  \Gamma(\phi \to \gamma\gamma) 
 &=& \frac{\alpha_{\rm EM}^2 M_\phi^3}{256 \pi^3 F_\phi^2} 
 \Bigg| 
 A_W(\tau_W) + 
 \sum_f  (3-\gamma_m)  N_c^{(f)} Q_f^2 A_f(\tau_f) 
  + 
2 b_F(e)
  \Bigg|^2 
  \,, \nonumber \\ 
 A_W(\tau_W) &=& - \left[ 2 + 3 \tau_W + 3 \tau_W (2-\tau_W) f(\tau_W) \right] 
 \,, \qquad 
 \tau_W = \frac{4 m_W^2}{M_\phi^2}
\,,  \nonumber \\ 
 A_{f}(\tau_{f}) &=&  2 \tau_{f} \left[ 1 + \left( 1 - \tau_{f} \right) f(\tau_{f}) \right] 
\,, 
\qquad 
\tau_{f} = \frac{4 m_{f}^2}{M_\phi^2}
\,, 
\nonumber \\ 
  f(\tau) 
&=&  
  \Bigg\{ 
  \begin{array}{cc} 
  \left( \sin^{-1} \frac{1}{\sqrt{\tau}} \right)^2 & {\rm for} \qquad \tau > 1 \nonumber \\ 
  - \frac{1}{4} \left[  \log \left( \frac{1 + \sqrt{1+\tau}}{1 - \sqrt{1-\tau}} \right) - i \pi \right] 
  & {\rm for} \qquad \tau \le 1 
  \end{array}
  \,, 
\nonumber \\ 
 b_F(e) 
&=&  \frac{(4\pi)^2 \beta_{F}^{\rm EM}(e)}{e^3}
= 
\frac{2}{3} N_{\rm TC} \sum_F N_c^{(F)} Q_F^2 
\,,
\label{TD:2gamma}
\end{eqnarray}
where $\alpha_{\rm EM}=e^2/(4\pi)$ and 
$N_c^{(F)}$ = 3(1) for techni-quarks (leptons); $Q_{f(F)}$ denotes the EM charge for 
SM $f$-fermions ($F$-techni-fermions).

\item $\phi \to gg$:

\begin{eqnarray} 
\Gamma(\phi \to gg) 
 &=& \frac{  \alpha_s^2 M_{\phi}^3}{32 \pi^3 F_{\phi}^2} 
 \Bigg|   
  \sum_{q} (3-\gamma_m) \tau_q \left[ 1 + \left( 1 - \tau_q \right) f(\tau_q) \right] 
  + b_F(g_s)  
  \Bigg|^2 
  \,, \nonumber \\ 
  b_F(g_s) 
&=& \frac{(4\pi)^2 \beta(g_s)}{g_s^3} 
  = \frac{2}{3} N_{\rm TC} \sum_Q N_Q
 \,, \label{TDgg} 
\end{eqnarray}
where $\alpha_s=g_s^2/(4\pi)$ with $g_s$ being $SU(3)_c$ gauge coupling, 
and $q$ and $Q$ denote SM quark and techni-quark, respectively.

\item $\phi \to WW, ZZ$:

\begin{eqnarray} 
\Gamma(\phi \to WW/ZZ) 
&= & \delta_{W(Z)} \, \frac{ M_{\phi}^3}{32 \pi F_{\phi}^2} 
\sqrt{1- \tau_{W/Z}} \left( 1 - \tau_{W/Z} + \frac{3}{4} \tau_{W/Z}^2 \right) 
\,, 
\qquad 
\delta_{W(Z)} =2(1)
\label{TD:WWZZ}
\end{eqnarray}

\end{itemize}

One can also incorporate higher order QCD corrections in the same way as done in the SM Higgs case~\cite{Spira:1997dg}, 
which would be relevant to $\phi \to $light quarks and $gg$ decay modes.


\begin{thebibliography}{99}

\bibitem{:2012si} 
  [ATLAS Collaboration],
arXiv:1202.1408 [hep-ex].  




\bibitem{ATLAS-CONF-2012-019}
The ATLAS collaboration, 
``{\it An update to the combined search for the Standard Model Higgs boson with the ATLAS detector at the LHC using 
up to 4.9 fb$^{-1}$ of $pp$ collision data at $\sqrt{s}=7$ TeV}", 
ATLAS-CONF-2012-019. 




\bibitem{Chatrchyan:2012tx} 
  S.~Chatrchyan {\it et al.}  [CMS Collaboration],
arXiv:1202.1488 [hep-ex].  


\bibitem{CMS-PAS-HIG-12-008}
The CMS collaboration, 
``{\it Combined results of searches for a Higgs boson in the context of the standard model and the beyond -standard models}", 
CMS-PAS-HIG-12-008. 



\bibitem{:2012sk} 
  [ATLAS Collaboration],
arXiv:1202.1414 [hep-ex].  




\bibitem{Chatrchyan:2012tw} 
  S.~Chatrchyan {\it et al.}  [CMS Collaboration],
arXiv:1202.1487 [hep-ex]. 


\bibitem{CMS-PAS-HIG-12-001} 
The CMS Collaboration, 
``{\it A search using multivariate techniques for a standard model Higgs boson decaying into two photons}" , 
CMS-PAS-HIG-12-001. 









    
  
\bibitem{Weinberg:1975gm}
  S.~Weinberg,
  Phys.\ Rev.\ D {\bf 13}, 974 (1976);
  L.~Susskind,
  Phys.\ Rev.\ D {\bf 20}, 2619 (1979).
 
 
 
\bibitem{Farhi:1980xs}
 For reviews, see, e.g.,  E.~Farhi and L.~Susskind,
  Phys.\ Rept.\  {\bf 74}, 277 (1981);
  K.~Yamawaki,
  Lecture at 14th  Symposium on Theoretical Physics, Cheju, Korea,  July 1995, 
  arXiv:hep-ph/9603293;
  C.~T.~Hill and E.~H.~Simmons,
  Phys.\ Rept.\  {\bf 381}, 235 (2003)
  [Erratum-ibid.\  {\bf 390}, 553 (2004)]; 
  F.~Sannino,
  Acta Phys.\ Polon.\  {\bf B40}, 3533-3743 (2009). 
  

\bibitem{Holdom:1981rm}
  B.~Holdom,
  Phys.\ Rev.\  D {\bf 24}, 1441 (1981).



  

\bibitem{Yamawaki:1985zg}
  K.~Yamawaki, M.~Bando and K.~Matumoto,
  Phys.\ Rev.\ Lett.\  {\bf 56}, 1335 (1986);
  M.~Bando, T.~Morozumi, H.~So and K.~Yamawaki,
  Phys.\ Rev.\ Lett.\ 
  {\bf 59}, 389 (1987).


\bibitem{Bando:1986bg}
  M.~Bando, K.~Matumoto and K.~Yamawaki,
  Phys.\ Lett.\  B {\bf 178}, 308 (1986). 



\bibitem{Lane:1991qh}
K.~D.~Lane and M.~V.~Ramana,
Phys. Rev. D {\bf 44}, 2678 (1991).  


\bibitem{Appelquist:1996dq}
T.~Appelquist, J.~Terning and L.~C.~Wijewardhana,
Phys. Rev. Lett. {\bf 77}, 1214 (1996); 
T.~Appelquist, A.~Ratnaweera, J.~Terning and L.~C.~Wijewardhana,
Phys. Rev. D {\bf 58}, 105017 (1998). 







\bibitem{Miransky:1996pd}
V.~A.~Miransky and K.~Yamawaki,
Phys. Rev. D {\bf 55}, 5051 (1997);
Errata, {\bf 56}, 3768 (1997). 





\bibitem{Caswell:1974gg}
  W.~E.~Caswell,
  Phys.\ Rev.\ Lett.\  {\bf 33}, 244 (1974); 
  T.~Banks and A.~Zaks,
  Nucl.\ Phys.\  B {\bf 196}, 189 (1982).



\bibitem{Appelquist:1991is}
  T.~Appelquist and G.~Triantaphyllou,
  Phys.\ Lett.\  B {\bf 278}, 345 (1992);
  R.~Sundrum and S.~D.~H.~Hsu,
  Nucl.\ Phys.\ B {\bf 391}, 127 (1993);
T.~Appelquist and F.~Sannino,
Phys.\ Rev.\ D {\bf 59}, 067702 (1999).


\bibitem{Harada:2005ru}
  M.~Harada, M.~Kurachi and K.~Yamawaki,
  Prog.\ Theor.\ Phys.\  {\bf 115}, 765 (2006);
  M.~Kurachi and R.~Shrock,
  Phys.\ Rev.\ D {\bf 74}, 056003 (2006);
  M.~Kurachi, R.~Shrock and K.~Yamawaki,
  Phys.\ Rev.\  D {\bf 76}, 036003 (2007).
  
\bibitem{Dimopoulos:1979es}
  S.~Dimopoulos and L.~Susskind,
  Nucl.\ Phys.\  B {\bf 155}, 237 (1979); 
  E.~Eichten and K.~D.~Lane,
  Phys.\ Lett.\  B {\bf 90}, 125 (1980).




\bibitem{Cacciapaglia:2004rb}
  G.~Cacciapaglia, C.~Csaki, C.~Grojean and J.~Terning,
  Phys.\ Rev.\  D {\bf 71}, 035015 (2005);
  R.~Foadi, S.~Gopalakrishna and C.~Schmidt,
  Phys.\ Lett.\  B {\bf 606}, 157 (2005); 
  R.~S.~Chivukula, E.~H.~Simmons, H.~J.~He, M.~Kurachi and M.~Tanabashi,
  Phys.\ Rev.\  D {\bf 72}, 016008 (2005). 




\bibitem{Miransky:1984ef}
  V.~A.~Miransky,
  Nuovo Cim.\  A {\bf 90}, 149 (1985).


\bibitem{Bardeen:1985sm} 
  W.~A.~Bardeen, C.~N.~Leung and S.~T.~Love,
  Phys.\ Rev.\ Lett.\  {\bf 56}, 1230 (1986); 
  Nucl.\ Phys.\ B {\bf 273}, 649 (1986).




\bibitem{Hashimoto:2010nw}
  M.~Hashimoto and K.~Yamawaki,
  Phys.\ Rev.\  D {\bf 83}, 016008 (2011). 
  
  
  
  
  
  


  \bibitem{Miransky:1989qc}
  V.~A.~Miransky and V.~P.~Gusynin,
  Prog. Theor. Phys. {\bf 81}, 426 (1989).
  





\bibitem{Haba:2010hu}
  K.~Haba, S.~Matsuzaki, K.~Yamawaki,
  Phys.\ Rev.\  {\bf D82}, 055007 (2010). 



\bibitem{Choi:2011fy} 
  K.~-Y.~Choi, D.~K.~Hong and S.~Matsuzaki,
Phys.\ Lett.\ B {\bf 706}, 183 (2011);   
 arXiv:1201.4988 [hep-ph].  






\bibitem{Yamawaki:2007zz} 
  K.~Yamawaki,
  Prog.\ Theor.\ Phys.\ Suppl.\  {\bf 167}, 127 (2007);
  Prog. Theor. Phys. Suppl. {\bf 180}, 1 (2010); 
  Int.\ J.\ Mod.\ Phys.\ A {\bf 25}, 5128 (2010). 






\bibitem{Shuto:1989te}
  S.~Shuto, M.~Tanabashi and K.~Yamawaki,
 in {\it Proc. 1989 Workshop on Dynamical Symmetry Breaking}, 
      Dec. 21-23, 1989, Nagoya, eds. T. Muta and K. Yamawaki 
      (Nagoya Univ., Nagoya, 1990) 115-123;    
%
%
%
  W.~A.~Bardeen, S.~T.~Love,
  Phys.\ Rev.\  {\bf D45}, 4672-4680 (1992); 
%
%
  M.~S.~Carena and C.~E.~M.~Wagner,
  Phys.\ Lett.\  B {\bf 285}, 277 (1992);
  M.~Hashimoto,
  Phys.\ Lett.\  B {\bf 441}, 389 (1998);   
%
%
  M.~Harada, M.~Kurachi and K.~Yamawaki,
  Phys.\ Rev.\ D {\bf 68}, 076001 (2003);
  M.~Kurachi and R.~Shrock,
  JHEP {\bf 0612}, 034 (2006). 





\bibitem{Kutasov:2011fr} 
  D.~Kutasov, J.~Lin and A.~Parnachev,
 Nucl.\ Phys.\ B {\bf 858}, 155 (2012).





\bibitem{Matsuzaki:2011ie} 
  S.~Matsuzaki and K.~Yamawaki,
Prog.\  Theor.\  Phys.\  {\bf 127}, , 209 (2012).   


\bibitem{Matsuzaki:2012gd} 
  S.~Matsuzaki and K.~Yamawaki,
  Phys.\ Rev.\ D {\bf 85}, 095020 (2012). 




\bibitem{Hashimoto:2011cw} 
  M.~Hashimoto,
Phys.\ Rev.\ D {\bf 84}, 111901 (2011).  





\bibitem{Jia:2012kd} 
  J.~Jia, S.~Matsuzaki and K.~Yamawaki,
  arXiv:1207.0735 [hep-ph].
  



\bibitem{Schechter:1980ak} 
  J.~Schechter,
  Phys.\ Rev.\ D {\bf 21}, 3393 (1980).







\bibitem{Pagels:1979hd}
  H.~Pagels and S.~Stokar,
  Phys.\ Rev.\  D {\bf 20}, 2947 (1979).



\bibitem{Appelquist:1988yc} 
  T.~Appelquist, K.~D.~Lane and U.~Mahanta,
  Phys.\ Rev.\ Lett.\  {\bf 61}, 1553 (1988).












\bibitem{Christensen:2005cb}
  N.~D.~Christensen and R.~Shrock,
  Phys.\ Lett.\  B {\bf 632}, 92 (2006); 
  M.~A.~Luty,
  JHEP {\bf 0904}, 050 (2009).












  
\bibitem{Georgi:1977gs}
  H.~M.~Georgi, S.~L.~Glashow, M.~E.~Machacek, D.~V.~Nanopoulos,
  Phys.\ Rev.\ Lett.\  {\bf 40}, 692 (1978).


\bibitem{Dittmaier:2011ti} 
  S.~Dittmaier {\it et al.}  [LHC Higgs Cross Section Working Group Collaboration],
arXiv:1101.0593 [hep-ph]; 
  LHC Higgs Cross Section Working Group {\it et al.},
  arXiv:1201.3084 [hep-ph].






\bibitem{Chatrchyan:2012ty}  
 S.~Chatrchyan {\it et al.}  [CMS Collaboration],
arXiv:1202.1489 [hep-ex];   
  S.~Chatrchyan {\it et al.}  [CMS Collaboration],
arXiv:1202.1997 [hep-ex]; 





\bibitem{:2012sm} 
  G.~Aad {\it et al.}  [ATLAS Collaboration],
arXiv:1202.1415 [hep-ex];   
The ATLAS Collaboration,
``{\it Search for the Standard Model Higgs boson in the $H \to WW^{(*)} \to l \nu l\nu$ decay mode with 4.7 fb$^{-1}$ of ATLAS data at $\sqrt{s}$ = 7 TeV}", 
ATLAS-CONF-2012-012. 







\bibitem{ATLAS-CONF-2012-014}
The ATLAS Collaboration, 
``{\it Search for the Standard Model Higgs boson in the $H \to \tau^+ \tau^-$ decay mode with 4.7 fb$^{-1}$ of ATLAS data at $\sqrt{s}$ = 7 TeV}", 
ATLAS-CONF-2012-014; 
The CMS Collaboration, 
``{\it Search for Neutral Higgs Bosons Decaying into Tau Leptons in the Di-Muon Channel in $pp$ Collisions at 7 TeV}", 
CMS-PAS-HIG-12-007. 


\bibitem{ATLAS-CONF-2012-015}
The ATLAS Collaboration, 
``{\it Search or the Standard Model Higgs boson produced in association with a vector boson and decaying to a $b$-quark pair using up to 
4.7 fb$^{-1}$ of $pp$ collision data at $\sqrt{s}$=7 TeV with the ATLAS detector at the LHC}", 
ATLAS-CONF-2012-015; 
  S.~Chatrchyan {\it et al.}  [CMS Collaboration],
arXiv:1202.4195 [hep-ex].  




\bibitem{TEVNPH:2012ab} 
  [TEVNPH (Tevatron New Phenomina and Higgs Working Group) and CDF and D0 Collaborations],
arXiv:1203.3774 [hep-ex].  




\bibitem{Azatov:2012bz} 
  A.~Azatov, R.~Contino and J.~Galloway,
arXiv:1202.3415 [hep-ph];   
  J.~R.~Espinosa, C.~Grojean, M.~Muhlleitner and M.~Trott,
arXiv:1202.3697 [hep-ph];   
  P.~P.~Giardino, K.~Kannike, M.~Raidal and A.~Strumia,
arXiv:1203.4254 [hep-ph];   
  T.~Li, X.~Wan, Y.~-k.~Wang and S.~-h.~Zhu,
arXiv:1203.5083 [hep-ph];  
  M.~Rauch,
arXiv:1203.6826 [hep-ph];   
  J.~Ellis and T.~You,
arXiv:1204.0464 [hep-ph].   




\bibitem{Campbell:2011iw} 
  B.~A.~Campbell, J.~Ellis and K.~A.~Olive,
JHEP {\bf 1203}, 026 (2012).   






\bibitem{Clark:1986gx} 
  T.~E.~Clark, C.~N.~Leung and S.~T.~Love,
  Phys.\ Rev.\ D {\bf 35}, 997 (1987).





 
  





\bibitem{Goldberger:2007zk}
  W.~D.~Goldberger, B.~Grinstein and W.~Skiba,
  Phys.\ Rev.\ Lett.\  {\bf 100}, 111802 (2008);  
  J.~Fan, W.~D.~Goldberger, A.~Ross and W.~Skiba,
  Phys.\ Rev.\  D {\bf 79}, 035017 (2009);  
  L.~Vecchi,
 Phys.\ Rev.\ D {\bf 82}, 076009 (2010);   
  B.~Coleppa, T.~Gregoire and H.~E.~Logan,
  arXiv:1111.3276 [hep-ph];
  V.~Barger, M.~Ishida and W.~-Y.~Keung,
  arXiv:1111.4473 [hep-ph]; 
  K.~Cheung and T.~-C.~Yuan,
  Phys.\ Rev.\ Lett.\  {\bf 108}, 141602 (2012)
  [arXiv:1112.4146 [hep-ph]].



\bibitem{070412} 
J.~Incandela, CMS talk at 
``{\it Latest update in the search for the Higgs boson at CERN}", 
July 4, 2012; 
F.~Gianotti, ATLAS talk at ``{\it Latest update in the search for the Higgs boson at
CERN}", July 4, 2012. 




\bibitem{Aoki:1990yp} 
  K.~-I.~Aoki, T.~Kugo and M.~G.~Mitchard,
  Phys.\ Lett.\ B {\bf 266}, 467 (1991).


\bibitem{Harada:2004qn} 
  M.~Harada, M.~Kurachi and K.~Yamawaki,
  Phys.\ Rev.\ D {\bf 70}, 033009 (2004). 
  



\bibitem{Miransky:1988gk}
  V.~A.~Miransky, K.~Yamawaki,
  Mod.\ Phys.\ Lett.\  {\bf A4}, 129-135 (1989).
  
  
  

\bibitem{Matumoto:1989hf}
  K.~Matumoto,
  Prog.\ Theor.\ Phys.\  {\bf 81}, 277-279 (1989);
%
  T.~Appelquist, M.~Einhorn, T.~Takeuchi, L.~C.~R.~Wijewardhana,
  Phys.\ Lett.\  {\bf B220}, 223 (1989).
  




\bibitem{Lawrance:2012cg} 
  R.~Lawrance and M.~Piai,
  arXiv:1207.0427 [hep-ph]; 
  R.~S.~Chivukula, B.~Coleppa, P.~Ittisamai, H.~E.~Logan, A.~Martin, J.~Ren and E.~H.~Simmons,
  arXiv:1207.0450 [hep-ph]; 
  I.~Low, J.~Lykken and G.~Shaughnessy,
  arXiv:1207.1093 [hep-ph]; 
  P.~P.~Giardino, K.~Kannike, M.~Raidal and A.~Strumia,
  arXiv:1207.1347 [hep-ph]; 
  J.~Ellis and T.~You,
  arXiv:1207.1693 [hep-ph]; 
  D.~Carmi, A.~Falkowski, E.~Kuflik, T.~Volansky and J.~Zupan,
  arXiv:1207.1718 [hep-ph]. 



  
  



\bibitem{Spira:1997dg}
  M.~Spira,
  Fortsch.\ Phys.\  {\bf 46}, 203 (1998)
  [arXiv:hep-ph/9705337].







\end{thebibliography}
\end{document}